\documentclass[twocolumn,aps,pr,superscriptaddress,preprintnumbers,nofootinbib,10pt]{revtex4-2}
\usepackage{amsmath}
\usepackage{amssymb}
\usepackage[dvipdf,dvips]{graphicx}
\usepackage{color}
\usepackage{hyperref}
\usepackage{url}
\usepackage{slashed}
\usepackage{subfigure}
\usepackage[usenames,dvipsnames]{xcolor}
\usepackage{amsmath}
\usepackage{amsfonts}
\usepackage{float} 
\usepackage{amssymb}
\usepackage{epsfig}
\usepackage{graphics}
\usepackage{euscript}
\usepackage{slashed}
\usepackage{epstopdf}
\usepackage[utf8]{inputenc}
\allowdisplaybreaks
\usepackage[normalem]{ulem}
\usepackage{pifont}
\usepackage{dsfont}
\usepackage{MnSymbol}
\usepackage{verbatim}
\usepackage{graphicx}
\usepackage{latexsym}
\usepackage{tikz-feynman}

\begin{document}
	
	\title{Probing Mermin's inequalities violations through pseudospin operators}

	\author{Philipe De Fabritiis} \email{pdf321@cbpf.br} \affiliation{CBPF $-$ Centro Brasileiro de Pesquisas Físicas, Rua Dr. Xavier Sigaud 150, 22290-180, Rio de Janeiro, Brazil}

	\author{Itzhak Roditi} \email{roditi@cbpf.br} \affiliation{CBPF $-$ Centro Brasileiro de Pesquisas Físicas, Rua Dr. Xavier Sigaud 150, 22290-180, Rio de Janeiro, Brazil}
	
	\author{Silvio P. Sorella} \email{silvio.sorella@gmail.com} \affiliation{UERJ $–$ Universidade do Estado do Rio de Janeiro, Rua São Francisco Xavier 524, 20550-013, Maracanã, Rio de Janeiro, Brazil}
	
	\begin{abstract}
	The violation of Mermin's inequalities is analyzed  by making use of two different Bell setups built with pseudospin operators. Employing entangled states defined by means of squeezed and coherent states,  the expectation value of Mermin's polynomials $M_n$ is evaluated for $n=3$ and $n=4$. In each case, we analyze the  correlator $\langle M_n \rangle$ and identify the set of parameters leading to the violation of Mermin's inequalities and to the saturation of the bound predicted by Quantum Mechanics.
	\end{abstract}

	\maketitle	
	
	\section{Introduction}
	\vspace{-0.3cm}
	
	Quantum entanglement can definitely be considered one of the most astounding aspects of Quantum Mechanics, and has been a topic of intensive research since its conception in the famous paper by Einstein, Podolski, and Rosen (EPR) in 1935~\cite{EPR35}. The abstract dilemma of a ``spooky action at a distance" introduced in the EPR paper was later reformulated in the form of concrete assumptions by John Bell in 1964~\cite{Bell64},  leading to falsifiable predictions that could be tested in the laboratory. Further, Clauser, Horne, Shimony, and Holt (CHSH) proposed a new version of Bell's inequality in 1969~\cite{CHSH89}, which was simpler and more suitable to be tested experimentally. Entangled particles exhibit quantum correlations that cannot be accounted for by any local realistic theory, and this intriguing feature can be revealed by the violation of Bell's inequalities, something that already had strong experimental evidence since 1982~\cite{Aspect82}, and has been robustly confirmed in many experiments over the last decades~\cite{Hensen15}. Nowadays, entanglement is considered a key resource for quantum information theory and  finds a huge number of applications~\cite{Zou21}. For a general review of Bell non-locality, one can see, for instance, Ref.~\cite{Brunner14}.

	The states examined in the original EPR paper had continuous degrees of freedom, and can be understood as a limiting case of the two-mode squeezed vacuum states. The preparation of these type of states for photons was investigated both theoretically~\cite{Reid88,Reid89,Walls94} and experimentally~\cite{Ou92a,Ou92b}. Furthermore, experiments reported the observation of Bell-type inequalities violation by the generalized EPR state produced in a pulsed nondegenerate optical parametric amplifier~\cite{Kuzmich00}, confirming previous theoretical predictions~\cite{Grangier88,Banaszek98a, Banaszek98b}. However, a more complete generalization of Bell's inequalities to continuous quantum variables only appeared in Ref.~\cite{Chen02}, where the authors have shown that two-mode squeezed vacuum states display quantum non-locality upon using generalized Bell's operators. Moreover, this work also showed that the original EPR states can maximally violate the Bell-CHSH inequality, saturating the Tsirelson's bound~\cite{Tsirelson80}. Interestingly enough, quantum systems with continuous variables find applications in some aspects of quantum information such as quantum computation~\cite{Lloyd99}, quantum teleportation~\cite{Vaidman94,Braunstein98a,Furusawa98}, among others~\cite{Braunstein98b,Lloyd98,Braunstein98c, Duan00, Cerf00}.
	
	It is possible to generalize quantum entanglement for systems with more than two subsystems~\cite{Greenberger07,Dur00,Briegel01}. Remarkably, multipartite entanglement is an important resource for quantum information theory~\cite{McCutcheon16}, and its experimental observation has been reported for the first time in Ref.~\cite{Bouwmeester99}. 
	There are generalizations of Bell inequalities for multipartite systems that were first proposed by Mermin~\cite{Mermin90}, and further developed by others~\cite{Ardehali92,Roy91,Belinskii93,Gisin98}. Mermin's inequalities are known to be maximally violated by GHZ-type states~\cite{GHZ90} and are also studied for W-type states~\cite{Swain18} (see also Refs.~\cite{Werner01,Cereceda01,Zukowski02}). The structure of correlations among more than two quantum systems was investigated in Ref.~\cite{Collins02b}, and Mermin's inequalities violations were reported with qubits made out of photons for 3-qubits in Refs.~\cite{Pan00,Erwen14} and 4-qubits in Ref.~\cite{Zhao03}. Furthermore, for superconducting qubits, a Bell-CHSH inequality violation was obtained in Ref.~\cite{Ansmann09}, and the violation of Mermin's inequality for 3-qubit in Refs.~\cite{DiCarlo10,Neeley10}. Quite recently, a loophole-free Bell inequality violation with superconducting circuits was demonstrated in Ref.~\cite{Storz23}. Let us also mention that Mermin's inequalities violations for 3, 4, and 5 superconducting qubits have been achieved using an IBM quantum computer in Ref.~\cite{Alsina16a}.  For an operational approach to build Bell's inequalities for multipartite systems, see Ref.~\cite{Alsina16b}. 

	It should be emphasized that in order to attest quantum non-locality, a given state does not need to violate all possible Bell's inequalities, it is enough that it violates {\it any} one of them~\cite{Chen02}. Therefore, although the quantum state is considered to be the source of entanglement and non-locality, the choice of Bell's operators has a crucial role in probing Bell's inequalities violation. 
	The strategy adopted in Ref.~\cite{Chen02} to test Bell's inequalities for bipartite systems with continuous variables was to consider pseudospin operators, providing an analogy between the two spin-$1/2$ particle system and the two-mode squeezed vacuum states, relying on the similar properties shared by pseudospin operators and Pauli matrices. The extension of these results to GHZ-type states~\cite{GHZ90} for multipartite quantum systems with continuous variables was done in Ref.~\cite{Zhang02}, and generalizations of the pseudospin operators were done in Refs.~\cite{Larsson03, Dorantes09}. Here, we will employ pseudospin operators in order to analyze Mermin's inequalities, adopting a Bell setup similar to the one used recently to investigate the Bell-CHSH inequality violation for entangled coherent states~\cite{BellCoherent}. For a formulation of Mermin's inequalities within the framework of relativistic Quantum Field Theory, see Ref.~\cite{MerminQFT}.
	
	The main goal of this work is to investigate Mermin's inequalities for multipartite systems considering entangled states different from the traditional GHZ-type states, namely, the squeezed-coherent and the squeezed-squeezed states. For each case, we analyze the set of parameters leading to the violation of Mermin's inequalities and, particularly, to the saturation of the Quantum Mechanics bound, providing thus examples of non-GHZ states which lead to a maximal violation. 
	
	The present work is organized as follows. In Section~\ref{SecMermin}, we introduce Mermin's polynomials and their corresponding inequalities. Pseudospin operators are used as building blocks for the Bell setups used  in Section~\ref{SecBell}. We analyze the violation of Mermin's inequalities for squeezed-coherent states in Section~\ref{SecSqueezCoher} and for squeezed-squeezed states in Section~\ref{SecSqueezSqueez}. Finally, in Section~\ref{SecConclusion}, we state our concluding remarks.

\vspace{-0.3cm}
\section{Mermin's inequalities}\label{SecMermin}
\vspace{-0.3cm}
Mermin's inequalities~\cite{Mermin90} can be used to test local realism for multipartite systems, generalizing Bell's inequalities from two-particle systems to a larger number of particles. The Mermin polynomials can be defined recursively through the general rule~\cite{Alsina16b,Collins02b}:
\begin{align}\label{key}
	M_n = \frac{1}{2} M_{n-1} \left(A_n + A'_n\right) + \frac{1}{2} M'_{n-1} \left(A_n - A'_n\right).
\end{align} 
In this formula, the $M_n'$ operators can be obtained from $M_n$ upon changing $A_n \rightarrow A_n'$ and $A_n' \rightarrow A_n$. We shall define the first term of this recursive procedure here by $M_1 = 2 A_1$. Here we are considering  the experimental situation in which $A_i$ and $A_i'$ are two dichotomic measurements that can be performed on each particle $i$ (with $i =1,... , n$), where the outcomes of these measurements can take the values $\pm1$.

Adopting this recursive construction for Mermin's polynomials, upon considering all the outcomes as $+1$, we obtain the classical upper bound for theories obeying local realism. On the other hand, in Quantum Mechanics, we are dealing with operators, requiring a detailed analysis of the possible eigenvalues of the corresponding Mermin operators. For instance, in the case  of sub-systems that only have two states, as qubits states,  the operators $A_i$ may be represented by $\vec{a}_i \cdot \vec{\sigma}$, where $\vec{\sigma} = (\sigma_1,\sigma_2,\sigma_3)$ are the Pauli matrices and $\vec{a}_i$ are three-vectors with unit norm. This leads to a Tsirelson-type bound~\cite{Alsina16b,Collins02b}:
\begin{align}\label{key}
	\vert M_n \vert \leq 2^{\left(\frac{n+1}{2}\right)}.
\end{align}	

In the sequence, we write a few  Mermin polynomials explicitly. The simplest one is given by:
\begin{align}\label{M2Def}
	M_2 = A B + A' B + A B'   - A' B'.
\end{align}
This is nothing more than the traditional Bell-CHSH operator. It is well-known that in absolute values, theories obeying local realism give a bound  $\vert\langle M_2 \rangle\vert_{\text{Cl}} \leq 2$, and Quantum Mechanics provides another bound, $\vert\langle M_2 \rangle\vert_{\text{QM}} \leq 2 \sqrt{2}$, the famous Tsirelson's bound~\cite{Tsirelson80}. Going further, we can write the Mermin polynomial for 3-qubits,
\begin{align}\label{M3Def}
	M_3 = A' B C + A B' C + A B C' - A' B' C'.
\end{align}
Considering the absolute value, we now have the classical bound $\vert\langle M_3 \rangle\vert_{\text{Cl}}  \leq 2 $. For   Quantum Mechanics, on the other hand, the upper bound is given by $\vert\langle M_3 \rangle\vert_{\text{QM}} \leq 4$. We remark that, upon choosing parameters such that $C = C' = 1$,  the operator $M_3$ reduces to $M_2$, that is, to the usual Bell-CHSH operator.
Finally, the Mermin polynomial for 4-qubits can be written as
\begin{align}\label{M4Def}
	2 M_4 &= -A B C D \nonumber \\
	&+ \left(A' B C D + A B' C D + A B C' D + A B C D'\right) \nonumber \\
	&+ \left( A' B' C D + A' B C' D + A' B C D'   \right. \nonumber \\
	&\left. \,\,\,   + A B' C' D + A B' C D' + A B C' D'      \right) \nonumber \\
	&-\left(A' B' C' D + A' B' C D' + A' B C' D' + A B' C' D'\right) \nonumber \\
	&-A' B' C' D'.
\end{align}
For this case, the classical bound is  $\vert\langle 2 M_4 \rangle\vert_{\text{Cl}} \leq 4$, while the quantum bound is given by $\vert\langle 2 M_4 \rangle\vert_{\text{QM}} \leq 8 \sqrt{2}$.
Interestingly, we remark that the violation of Mermin's inequalities for $M_3$, $M_4$, and $M_5$ have been verified using an IBM quantum computer in Ref.~\cite{Alsina16a}.

In order to investigate Mermin's inequalities, it is very common in the literature to use GHZ-type states (see for instance Refs.~\cite{Alsina16a,Zhang02,MerminQFT}). Here we will follow a different route and use entangled states built with coherent and squeezed states to search for violations of Mermin's inequalities corresponding to $M_3$ and $M_4$. One can define coherent states as annihilation operator eigenstates that can be represented in the Fock basis as
\begin{align}\label{Coherent}
	\vert \alpha \rangle = e^{-\frac{\alpha^2}{2}} \sum_{n=0}^{\infty} \frac{\alpha^n}{\sqrt{n!}} \, \vert n \rangle \;,
\end{align}
where $\alpha$ is taken as a real number, and we define the Fock basis states as $\vert n \rangle = \frac{(a^\dagger)^n}{\sqrt{n!}} \vert 0 \rangle$. Notice that, by construction, we have the property $a \vert \alpha \rangle = \alpha \vert \alpha \rangle$. Entangled coherent states are extremely interesting and have aroused much interest in the literature, as one can see in Ref.~\cite{BellCoherent} and references therein.
Furthermore, we can also define the squeezed states using the following expression:
\begin{align}\label{Squeezed}
	\vert \eta \rangle_{a b} = \sqrt{1 - \eta^2} \sum_{n=0}^{\infty} \eta^n \, \vert n_a, n_b \rangle,
\end{align}
where we defined $\vert x_a, y_b \rangle \equiv \vert x \rangle_a \otimes \vert y \rangle_b$ and $\eta$ is the so-called squeezing parameter, such that $\eta \in (0,1)$.  It is noteworthy that the two-mode squeezed vacuum state discussed in the introduction is of this type and can be understood in some sense as a regularized version of the original EPR state, as we have already highlighted.

In the next section, we discuss the Bell setup used in this work in order to study Mermin's inequalities. We remark that the idea of a pairing mechanism in the Hilbert space goes back to Ref.~\cite{Gisin92} and was recently discussed in the context of Bell-CHSH inequalities in Refs.~\cite{Peruzzo23, Sorella23a, Sorella23b}.

\section{Building the Bell setup with pseudospin operators}\label{SecBell}
	
The Bell operator for two-qubit systems (for example, spin-$1/2$ systems) can be written as
\begin{align}\label{key}
		\mathcal{B} &= \left( \vec{a} \cdot \vec{\sigma_1}\right) \otimes \left( \vec{b} \cdot \vec{\sigma_2}\right) + \left( \vec{a} \cdot \vec{\sigma_1}\right) \otimes \left( \vec{b'} \cdot \vec{\sigma_2}\right) \nonumber \\
		&+ \left( \vec{a'} \cdot \vec{\sigma_1}\right) \otimes \left( \vec{b} \cdot \vec{\sigma_2}\right) - \left( \vec{a'} \cdot \vec{\sigma_1}\right) \otimes \left( \vec{b'} \cdot \vec{\sigma_2}\right),
\end{align}
where $\{\vec{a}, \vec{a'}, \vec{b}, \vec{b'}\}$ are three-vectors with unit norm and  $\sigma_i$ is the Pauli matrix associated with the $i$-th qubit (here, $i=1,2$). From this definition, it is possible to obtain 
\begin{align}\label{key}
			\mathcal{B}^2 = 4 \mathbf{1}_{2 \times 2} + 4 \left[\left(\vec{a} \times \vec{a'}\right) \cdot \vec{\sigma_1}\right] \otimes \left[\left(\vec{b} \times \vec{b'}\right) \cdot \vec{\sigma_1}\right],
\end{align}
where $\mathbf{1}_{2 \times 2}$ is the identity matrix for this two-qubit system. Thus, from the above expression, one can find an upper bound for the expectation value of $\mathcal{B}^2$ with respect to a two-qubit state: $\langle \mathcal{B}^2 \rangle \leq 8$. Therefore, we have $\vert \langle \mathcal{B} \rangle \vert \leq 2 \sqrt{2} $, the Tsirelson's bound~\cite{Tsirelson80} cited before.

A generalization of spin operators for infinite-dimensional systems can be done using the so-called {\it pseudospin operators}~\cite{Chen02, Dorantes09,Larsson03}, that can be defined in the Fock basis through the following expressions:
\begin{align}\label{pseudospin}
		s_x &= \sum_{n=0}^{\infty} \vert 2n+1 \rangle \langle 2n \vert + \vert 2n \rangle \langle 2n + 1 \vert, \nonumber \\
		s_y &= i \sum_{n=0}^{\infty} \vert 2n \rangle \langle 2n + 1 \vert - \vert 2n + 1 \rangle \langle 2n \vert, \nonumber \\
		s_z &= \sum_{n=0}^{\infty} \vert 2n+1 \rangle \langle 2n + 1 \vert - \vert 2n \rangle \langle 2n \vert.
\end{align}
The pseudospin operators defined above have the same properties as the Pauli matrices, obeying identical commutation relations as those of the spin-$1/2$ systems, {\it i.e.},
\begin{align} 
	\left[ s_x,s_y \right] = 2 i s_z, \quad 	\left[s_y,s_z \right] = 2 i s_x, \quad 	\left[s_z,s_x \right] = 2 i s_y.
\end{align} 
Therefore, the pseudospin operator $\vec{s} = \left(s_x, s_y, s_z\right)$ can be regarded as a counterpart of the spin operator $\vec{\sigma}$. Considering a three-vector $\vec{a}$ with unit norm, we can immediately find $\left(\vec{a} \cdot \vec{s}\right)^2 = 1$. Consequently, the outcome of the Hermitian operator $\vec{a} \cdot \vec{s}$ measurement is $\pm1$, allowing us to interpret $\vec{a}$ as being the direction along which we measure the pseudospin $\vec{s}$, providing a complete analogy with the spin-$1/2$ case, as emphasized in Ref.~\cite{Chen02}. In particular, we can rewrite the Bell-CHSH inequality by simply substituting the spin operators $\sigma_i$ by their corresponding pseudospin operators $s_i$ and find totally analogous results, including the bounds for local realistic theories and the Tsirel'son bound, but now for quantum systems described by continuous variables.
	
In order to investigate the Mermin inequality corresponding to the polynomial $M_n$, we will consider a set of $n$ Bell operators, whose action on the states will be defined by two different Bell setups. In the first setup, the Bell operator $A_i$ will act in the $i$-th entry as
\begin{align}\label{key}
	A_i  &\equiv (\cos a_i) s_x^{(0)} - (\sin a_i) s_y^{(0)} + \sum_{n=2}^{\infty} \vert n \rangle \langle n \vert
\end{align}
where	$s_x^{(0)} \equiv \vert 1 \rangle \langle 0 \vert + \vert 0 \rangle \langle 1 \vert$ and $s_y^{(0)} \equiv i \left( \vert 0 \rangle \langle 1 \vert - \vert 1 \rangle \langle 0 \vert\right)$ are the first contributions of the infinite  sums corresponding to $s_x$ and $s_y$, respectively.
In other words, the Bell operators action on the Hilbert space can be defined by: 
\begin{align}\label{BellSetup1}
	A_i \vert ..., x_{i-1}, 0, x_{i+1}, ... \rangle &= e^{i a_i} \vert ..., x_{i-1}, 1, x_{i+1}, ... \rangle \nonumber \\
	A_i \vert  ..., x_{i-1}, 1, x_{i+1}, ... \rangle &= e^{-i a_i} \vert ..., x_{i-1}, 0, x_{i+1}, ... \rangle,
\end{align}
where $a_i$ are arbitrary real parameters, which can be chosen at the best convenience. In the second setup, the Bell operators can be written in term of pseudospin operators acting on the $i$-th space as
	\begin{align}\label{key}
		A_i \equiv (\cos a_i) \, s_x - (\sin a_i) \, s_y,
	\end{align}
	such that, its action on the Hilbert space is given by
	\begin{align}\label{BellSetup2}
	 \!\!\! A_i \vert ..., x_{i-1}, 2n, x_{i+1}, ... \rangle \!&=\! e^{ia_i} \vert  ..., x_{i-1}, 2n \! + \! 1, x_{i+1}, ... \rangle \nonumber \\
	 \!\!\!\! A_i \vert ..., x_{i-1}, 2n \! + \! 1, x_{i+1}, ... \rangle \!&=\! e^{-ia_i} \vert ..., x_{i-1}, 2n, x_{i+1}, ... \rangle,
	\end{align}
where $a_i$ are arbitrary real parameters. We remark that in both cases, the only non-trivial action of the operator $A_i$ happens in the $i$-th entry of the corresponding state, and the action in any other state is given by the identity. Notice that with these definitions,  we immediately achieve the main properties that we need in order to analyze Mermin's inequalities violation in both Bell setups, that is, these Bell operators $A_i$ satisfy:
	\begin{align}\label{BellProp}
		A_i^2 = 1, \quad A_i^\dagger = A_i, \quad \left[A_i, A_j\right] = 0.
	\end{align}

Now we are ready to study Mermin's inequalities through these Bell setups built with pseudospin operators~\eqref{pseudospin}. More precisely,  we shall investigate Mermin's inequality violation for $n$-partite systems using {\it squeezed-coherent states} when we have $n=3$ and {\it squeezed-squeezed states} when we have $n=4$. In each case, we compute the expectation value of the product $A_1 .... A_n$ in the corresponding state of interest, and use this result in order to find the correlator $\langle M_n \rangle$. Then, we analyze the set of parameters leading to a violation of Mermin's inequalities, and also investigate the possibility of saturating the bound predicted by Quantum Mechanics, finding the maximal violation. Finally, it is worth pointing out that in principle we could generalize the strategy adopted here for any $n$, employing squeezed-coherent and squeezed-squeezed states for $n$ odd and even, respectively.

	\section{Squeezed-Coherent states}\label{SecSqueezCoher}
	\vspace{-0.3cm}
	
	Let us begin with the investigation of Mermin's inequality violation for 3-partite systems. In order to do so, we define the {\it squeezed-coherent} entangled state:
	\begin{align}\label{Squeezed-Coherent}
		\vert \psi \rangle_{SC} = N_{SC} \left( \vert \eta \rangle_{ab} \vert \alpha \rangle_c + e^{i \phi} \vert \alpha \rangle_{ab} \vert \eta \rangle_c \right),
	\end{align}
	where $\vert \eta \rangle_{ab}$ and $\vert \alpha \rangle_{ab}$ represent squeezed states belonging to the product Hilbert space $\mathcal{H}_a \otimes \mathcal{H}_b$ as defined in Eq.~\eqref{Squeezed}, with squeezing parameters $\eta$ and $\alpha$, respectively. Besides, $\vert \alpha \rangle_c$ and $\vert \eta \rangle_c$ represent coherent states belonging to the Hilbert space $\mathcal{H}_c$ as defined in Eq.~\eqref{Coherent}, with parameters $\alpha$ and $\eta$, respectively.
	The normalization of this entangled squeezed-coherent state~\eqref{Squeezed-Coherent} is given by
	\begin{align}\label{OmegaSC}
		N_{SC} = \frac{1}{\sqrt{2}} \left[1 + \cos \phi \,  \frac{\sqrt{1-\alpha^2} \sqrt{1-\eta^2}}{1 - \alpha \eta} e^{-\frac{1}{2}(\alpha-\eta)^2}\right]^{-\frac{1}{2}}\!\!.
	\end{align}
	
	In the following, we will analyze the Mermin's inequality corresponding to the polynomial $M_3$, adopting the Bell setups defined by Eqs.~\eqref{BellSetup1},\eqref{BellSetup2}. We remark that in both cases, it is sufficient to compute the correlator $\langle A B C \rangle$, since for the primed operators it is sufficient to substitute the parameters $a_i$ by their primed version $a_i'$. Thus, with the $\langle A B C \rangle$ expression, we can immediately use Eq.~\eqref{M3Def} to construct the correlator $\langle M_3 \rangle$, whose full expression will be omitted here to not clutter the text.

		\begin{figure}[t!]
		\begin{minipage}[b]{1.0\linewidth}
			\includegraphics[width=\textwidth]{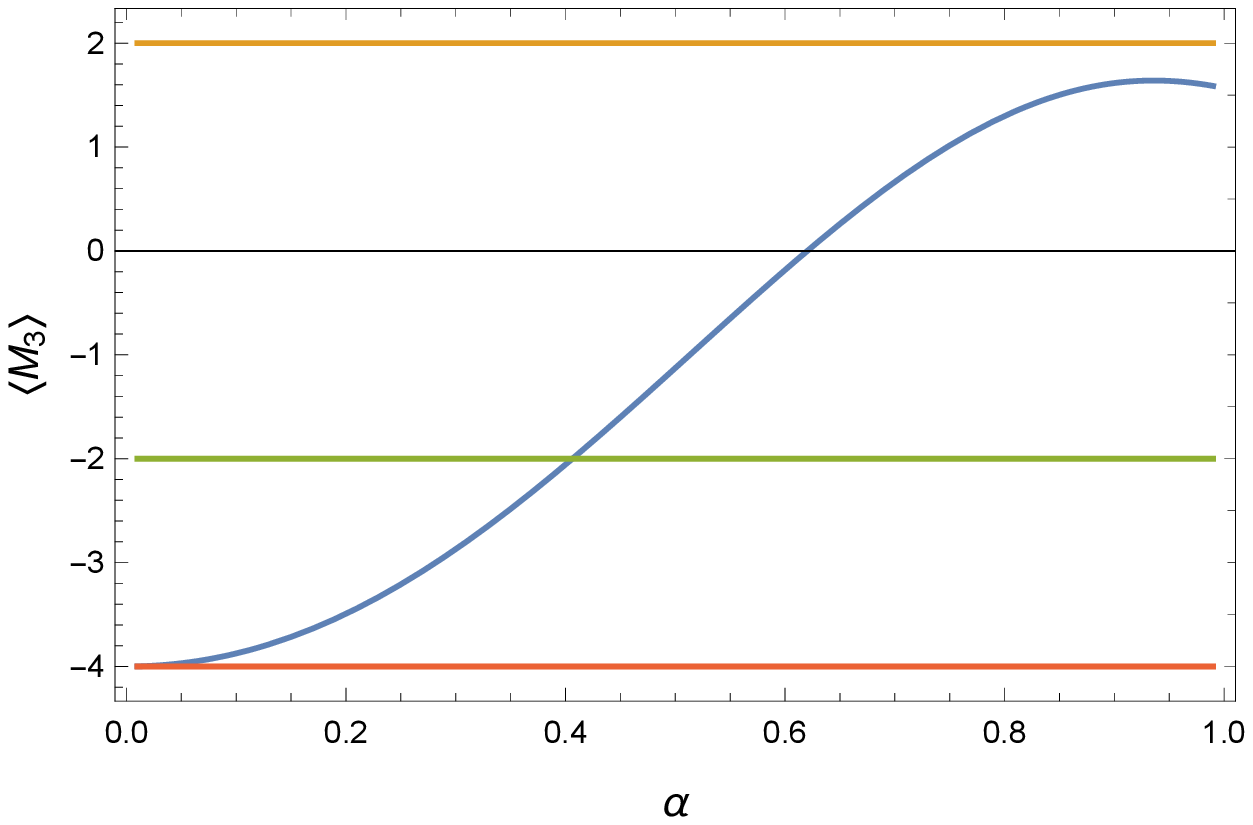}
		\end{minipage} \hfill
		\begin{minipage}[b]{1.0\linewidth}
			\includegraphics[width=\textwidth]{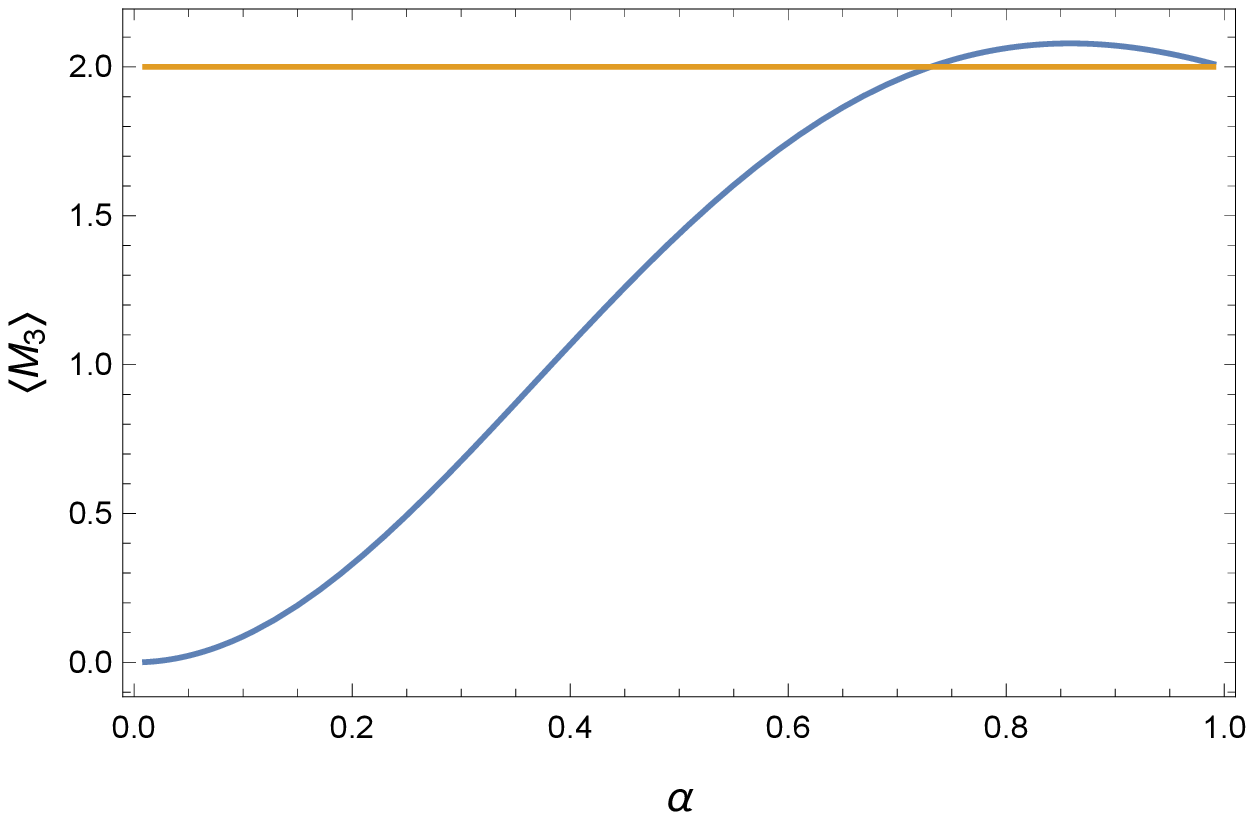}
		\end{minipage} \hfill
		\caption{$\langle M_3 \rangle$ as a function of $\alpha$ with $\eta = \alpha + 0.001$  in the first setup~\eqref{BellSetup1} for the squeezed-coherent state~\eqref{Squeezed-Coherent}. In the upper panel, we adopted $\phi=\pi$ and $a=0, \, a'=\pi/2, \, b= - b' = -\pi/4, \, c= - c' = +\pi/4$; in the lower panel, we have $\phi=0$ and $a=-\pi/4, \, a'=\pi/2, \, b= - b' = -\pi/4, \, c=0, \, c'=+\pi/4$. There is violation whenever we have $\vert \langle M_3 \rangle \vert >2$. The red line represents the Quantum Mechanics lower bound, given by $\langle M_3 \rangle = -4$.}
		\label{M3scTwo2D}
	\end{figure}
\begin{figure}[t!]
	\begin{minipage}[b]{1.0\linewidth}
		\includegraphics[width=\textwidth]{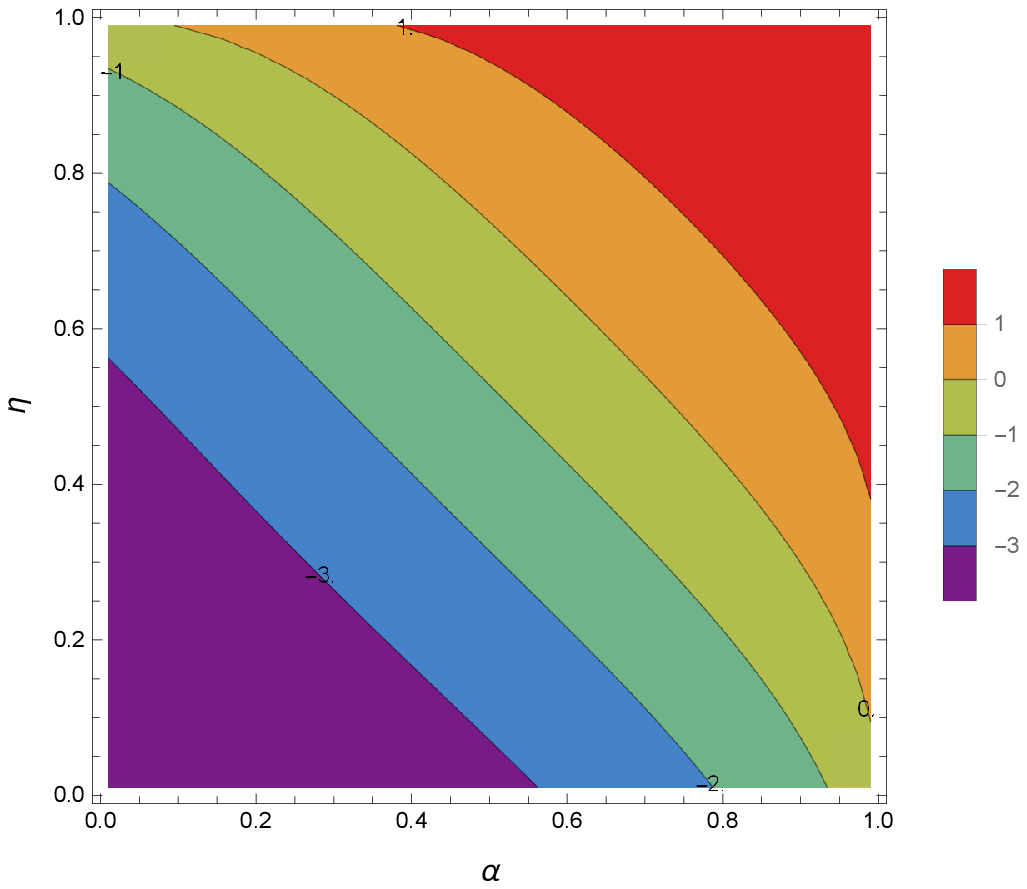}
	\end{minipage} \hfill
	\begin{minipage}[b]{1.0\linewidth}
	\includegraphics[width=\textwidth]{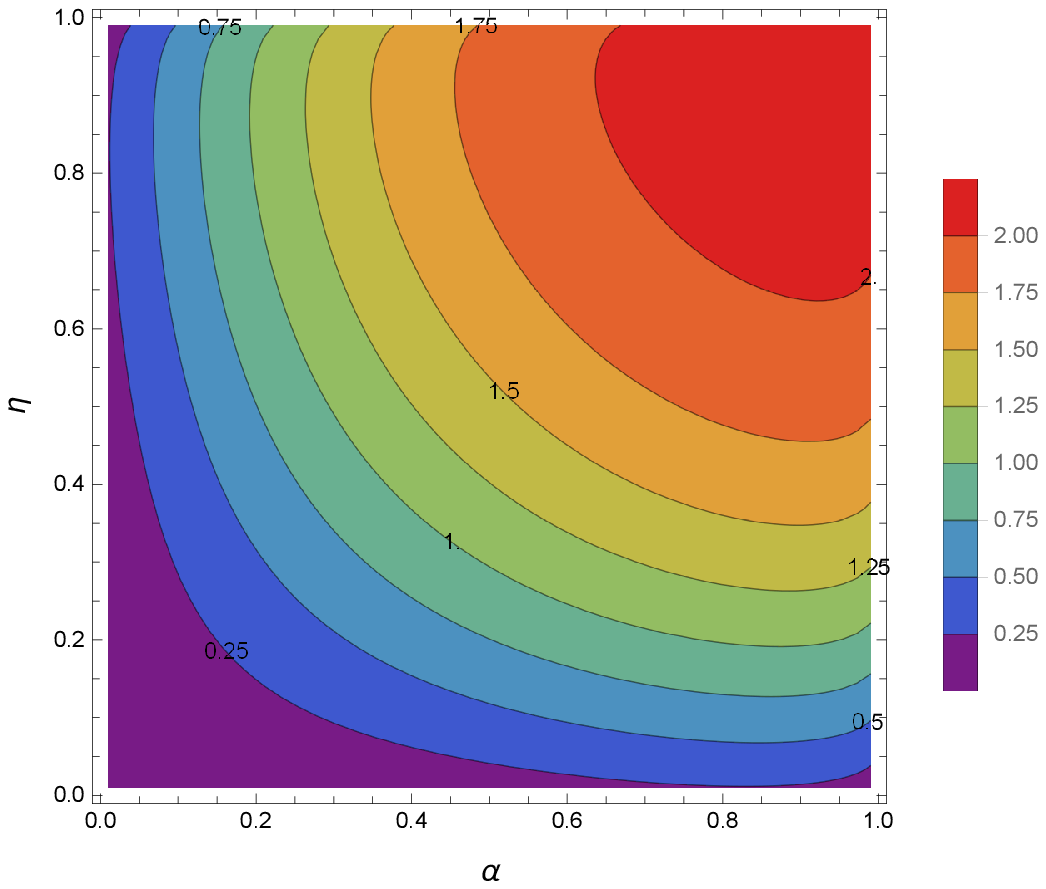}
\end{minipage} \hfill
	\caption{$\langle M_3 \rangle$ as a function of $\alpha$ and $\eta$ in the first setup~\eqref{BellSetup1} for the squeezed-coherent state~\eqref{Squeezed-Coherent}. In the upper panel, we adopted $\phi=\pi$ and $a=0, \, a'=\pi/2, \, b= - b' = -\pi/4, \, c= - c' = +\pi/4$; in the lower panel, we have $\phi=0$ and $a=-\pi/4, \, a'=\pi/2, \, b= - b' = -\pi/4, \, c=0, \, c'=+\pi/4$. There is violation whenever we have $\vert \langle M_3 \rangle \vert >2$.}
	\label{M3scTwoContour}
\end{figure}



	\vspace{-0.2cm}
	\subsection{First setup}
	\vspace{-0.2cm}
	First, we compute the correlator $\langle A B C \rangle$ considering the entangled squeezed-coherent state~\eqref{Squeezed-Coherent}, adopting the first Bell setup defined by Eq.~\eqref{BellSetup1}. Thus, we find:
	\begin{align}\label{key}
		&\langle ABC \rangle = N_{SC}^2 \Big\{ 4 \alpha \eta \left(K_1^2 + K_2^2\right) \cos(a+b) \cos c \nonumber \\
		&+ 4 \alpha \eta K_1 K_2 \cos \phi \cos(a+b+c) \nonumber \\
		&+ 2 K_1 K_2 \big[\eta^2 \cos(a+b-c+\phi) + \alpha^2 \cos(a+b-c-\phi)\big] \nonumber \\
		&+2 \cos (a+b) \big[ \eta K_1^2 \left(e^{\alpha^2} - 1 - \alpha^2\right) + \alpha K_2^2  \left(e^{\eta^2} - 1 - \eta^2\right)\big] \nonumber \\
		&+2 K_1 \! K_2  \left(e^{\alpha \eta} \!-\! 1 \!-\! \alpha \eta\right) \big[\eta \cos(a+b+\phi) \!+\! \alpha \cos(a+b-\phi)\big] \nonumber \\
		&+2 \cos c \big[\alpha K_1^2 \left(\frac{1}{1-\eta^2} - 1 - \eta^2\right) \!+\! \eta K_2^2 \left(\frac{1}{1-\alpha^2} - 1 - \alpha^2\right)\big] \nonumber \\
		&+2 K_1 K_2 \left(\frac{1}{1-\alpha \eta} - 1 - \alpha \eta\right) \big[\alpha \cos(c+\phi) + \eta \cos(c-\phi)\big] \nonumber \\
		&+K_1^2 \left(\frac{1}{1-\eta^2} - 1 - \eta^2\right)\left(e^{\alpha^2} - 1 - \alpha^2\right) \nonumber \\
		&+ K_2^2 \left(\frac{1}{1-\alpha^2} - 1 - \alpha^2\right)\left(e^{\eta^2} - 1 - \eta^2\right)  \nonumber \\
		&+2 \cos\phi K_1 K_2 \left(\frac{1}{1-\alpha \eta} - 1  -\alpha \eta\right)\left(e^{\alpha \eta} - 1 - \alpha \eta\right) \Big\},
	\end{align}
	where $K_1 = \sqrt{1-\eta^2} \, e^{-\alpha^2/2}$, $K_2 = \sqrt{1-\alpha^2} \, e^{-\eta^2/2}$, and the overall factor $N_{SC}$ is defined in Eq.~\eqref{OmegaSC}. With this expression at hand, we can immediately write the correlator $\langle M_3 \rangle$ adopting the definition presented in Eq.~\eqref{M3Def}, as we have already highlighted, and proceed to the numerical analysis, searching for a violation of the corresponding Mermin's inequality, that is, trying to find a set of parameters such that $\vert \langle M_3 \rangle \vert > 2$. 
	
	We were able to find a violation by choosing a relative phase $\phi = \pi$ and adopting the following set of parameters: $a=0, \, a'=\pi/2, \, b=-\pi/4, \, b' = +\pi/4, \, c=+\pi/4, \, c'=-\pi/4$. In fact, with these parameters, we obtain $\vert \langle M_3 \rangle \vert = 2.0163$ if we take $(\alpha, \eta) = (0.40, 0.41)$. Considering  $(\alpha, \eta) = (0.1, 0.7)$ we find $\vert \langle M_3 \rangle \vert = 2.0559$, thus it is not necessary to keep them close to each other to have a violation. The Quantum Mechanics bound $\vert \langle M_3 \rangle \vert = 4$ is saturated for $(\alpha, \eta) = (0.0010, 0.0011)$. 
	We have also found a small violation by considering $\phi = 0$, with the set of parameters: $a=-\pi/4, \, a'=\pi/2, \, b=-\pi/4, \, b' = +\pi/4, \, c=0, \, c'=+\pi/4$. In this case we find, for instance, $\vert \langle M_3 \rangle \vert = 2.0698$ if we take $(\alpha, \eta) = (0.90, 0.91)$, and $\vert \langle M_3 \rangle \vert = 2.0428$ if we take $(\alpha, \eta) = (0.7, 0.9)$. However, here we didn't find any violation taking small values for $\alpha$ and $\eta$, nor considering these parameters far from each other.
	
	We exhibit the value of $ \langle M_3 \rangle $ as a function of $\alpha$, considering $\eta = \alpha + 0.001$ in Fig.~\ref{M3scTwo2D}. Notice that for $\phi=\pi$, there is violation only for small values of $\alpha$ and the Quantum Mechanics bound is saturated when we take very small values of $\alpha$. On the other hand, for $\phi=0$, there is only a small violation,  occurring for big values of $\alpha$. The complete region of parameters $\alpha$ and $\eta$ leading to the violation of $\vert \langle M_3 \rangle \vert \leq 2$ is shown in Fig.~\ref{M3scTwoContour}.

	\vspace{-0.2cm}
	\subsection{Second setup}
	\vspace{-0.2cm}
	Now, continuing our analysis, we consider the second setup described by Eq.~\eqref{BellSetup2}. Thus, the correlator $\langle ABC \rangle$ in the squeezed-coherent state~\eqref{Squeezed-Coherent} is given by
	\begin{align}\label{key}
			&\langle ABC \rangle = N_{SC}^2 \sum_{m=0}^{\infty} \frac{1}{\sqrt{(2m !)} \sqrt{(2m +1 !)}} \Bigg\{  \nonumber \\
			&+ 4 \alpha \eta \cos(a+b) \cos c  \left(\frac{K_1^2 \alpha^{4m}}{1-\eta^4} + \frac{K_2^2 \eta^{4m}}{1-\alpha^4}\right) \nonumber \\
			&+ \frac{2 K_1 K_2 (\alpha \eta)^{2m}}{1-\alpha^2 \eta^2} \big[ 2 \cos \phi \cos(a+b+c) \alpha \eta \nonumber \\
			&+\eta^2 \cos(a+b-c+\phi) + \alpha^2 \cos(a+b-c-\phi) \big] \Bigg\},
	\end{align}
	where again $K_1 = \sqrt{1-\eta^2} \, e^{-\alpha^2/2}$, $K_2 = \sqrt{1-\alpha^2} \, e^{-\eta^2/2}$, and the overall $N_{SC}$ is defined in Eq.~\eqref{OmegaSC}. Here, unfortunately, we were not able to find a closed analytic expression for the infinite sum.  Even though, we can perform the analysis by considering the necessary number of terms in the series to satisfy the required precision. 
	
	Once again, we were able to find a violation with a relative phase $\phi = \pi$ and the same set of parameters: $a=0, \, a'=\pi/2, \, b=-\pi/4, \, b' = +\pi/4, \, c=+\pi/4, \, c'=-\pi/4$. Here we obtain $\vert \langle M_3 \rangle \vert = 2.0687$ if we take $(\alpha, \eta) = (0.4, 0.41)$ and $\vert \langle M_3 \rangle \vert = 2.0617$ considering $(\alpha, \eta) = (0.1, 0.7)$.  The Quantum Mechanics bound $\vert \langle M_3 \rangle \vert = 4$ is saturated for $(\alpha, \eta) = (0.0010, 0.0011)$. We have also found a small violation without a relative phase ($\phi=0$) with parameters $a=-\pi/4, \, a'=\pi/2, \, b=-\pi/4, \, b' = +\pi/4, \, c=0, \, c'=+\pi/4$. Here we find $\vert \langle M_3 \rangle \vert = 2.1067$ if we take $(\alpha, \eta) = (0.9, 0.91)$ and $\vert \langle M_3 \rangle \vert = 2.0216$ considering $(\alpha, \eta) = (0.7, 0.9)$. 
	
\begin{figure}[t!]
	\begin{minipage}[b]{1.0\linewidth}
		\includegraphics[width=\textwidth]{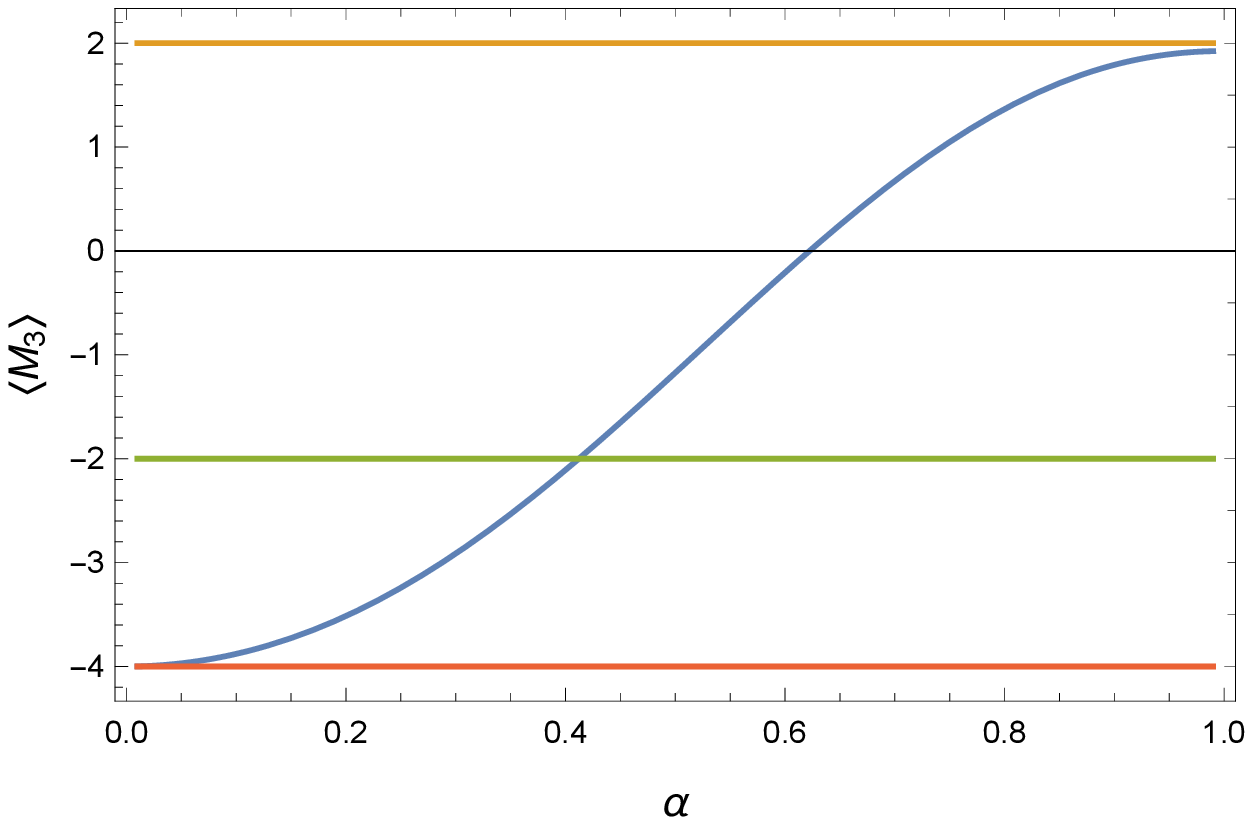}
	\end{minipage} \hfill
	\begin{minipage}[b]{1.0\linewidth}
		\includegraphics[width=\textwidth]{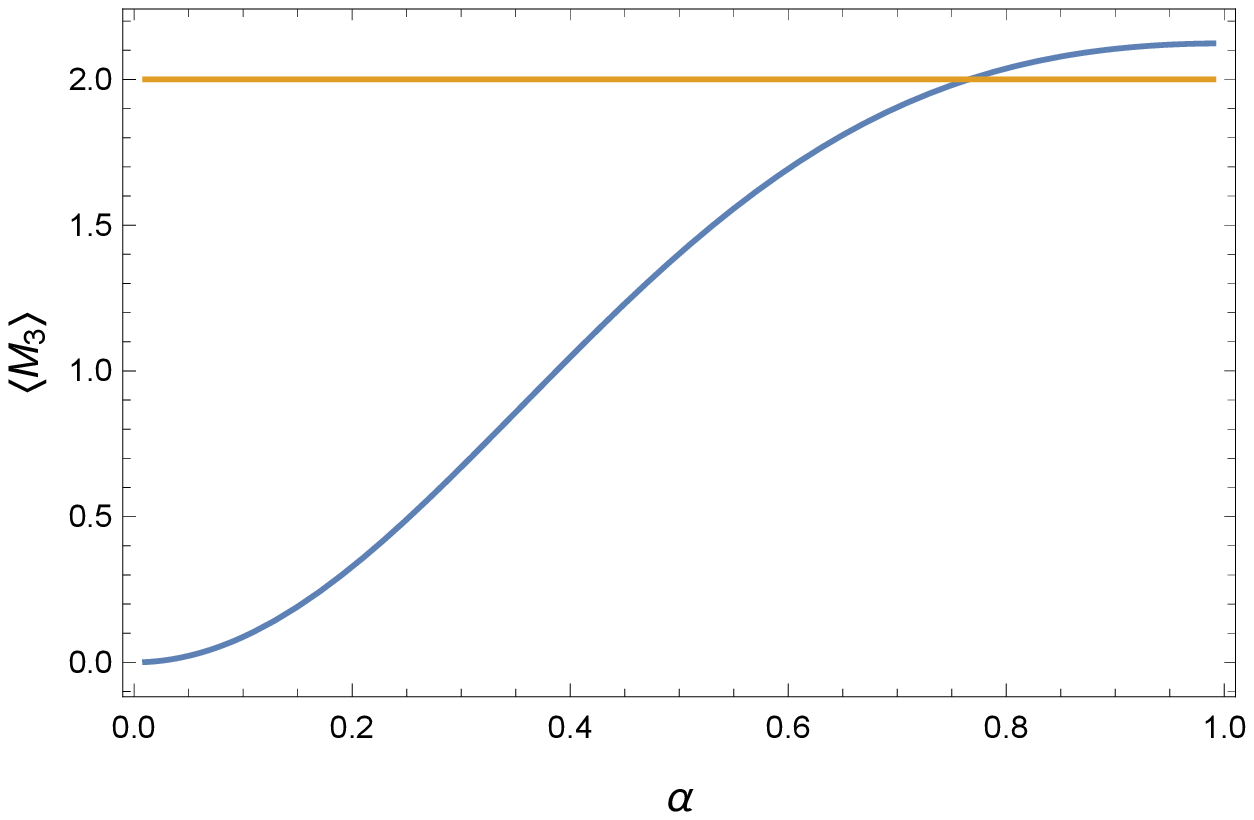}
	\end{minipage} \hfill
	\caption{$\langle M_3 \rangle$ as a function of $\alpha$ with $\eta = \alpha + 0.001$ in the second setup~\eqref{BellSetup2} for the squeezed-coherent state~\eqref{Squeezed-Coherent}. In the upper panel, we adopted $\phi=\pi$ and $a=0, \, a'=\pi/2, \, b= - b' = -\pi/4, \, c= - c' = +\pi/4$; in the lower panel, we have $\phi=0$ and $a=0, \, a'=\pi/2, \, b= -b' = -\pi/4, \, c= - c' = +\pi/4$. There is violation whenever we have $\vert \langle M_3 \rangle \vert >2$. The red line represents the Quantum Mechanics lower bound, given by $\langle M_3 \rangle = -4$.}
	\label{M3scAll2D}
\end{figure}

\begin{figure}[t!]
	\begin{minipage}[b]{1.0\linewidth}
		\includegraphics[width=\textwidth]{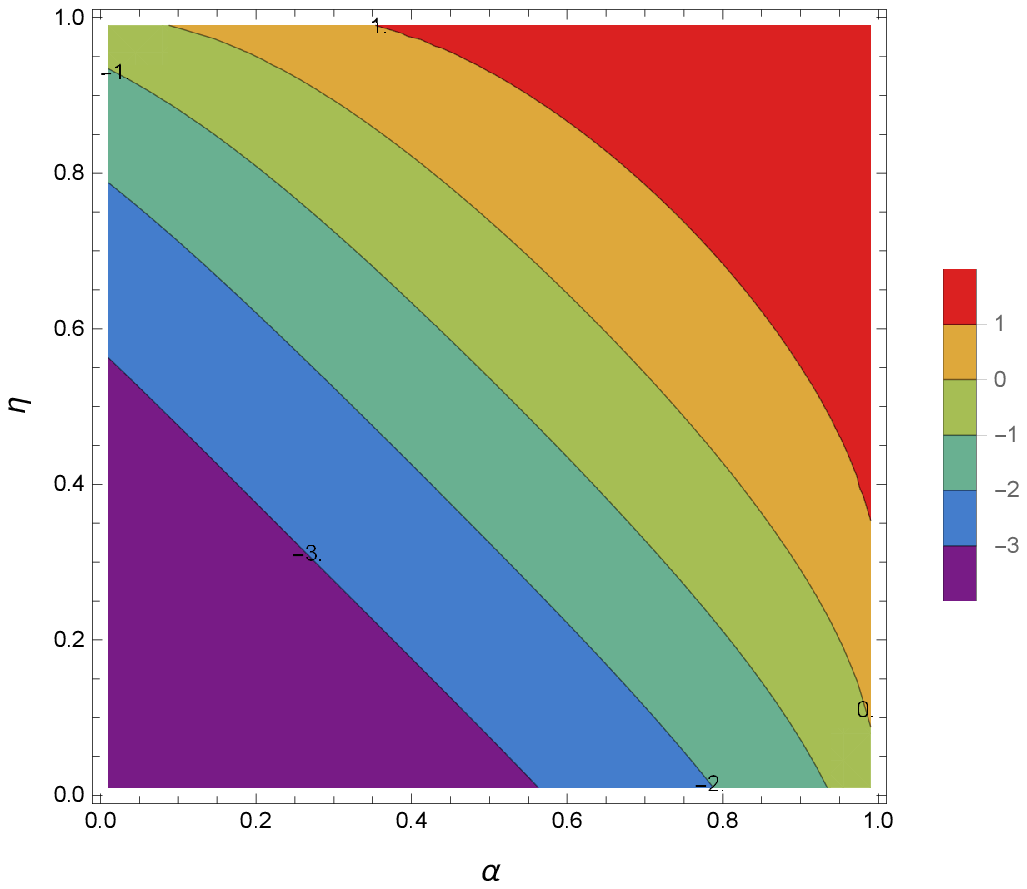}
	\end{minipage} \hfill
	\begin{minipage}[b]{1.0\linewidth}		\includegraphics[width=\textwidth]{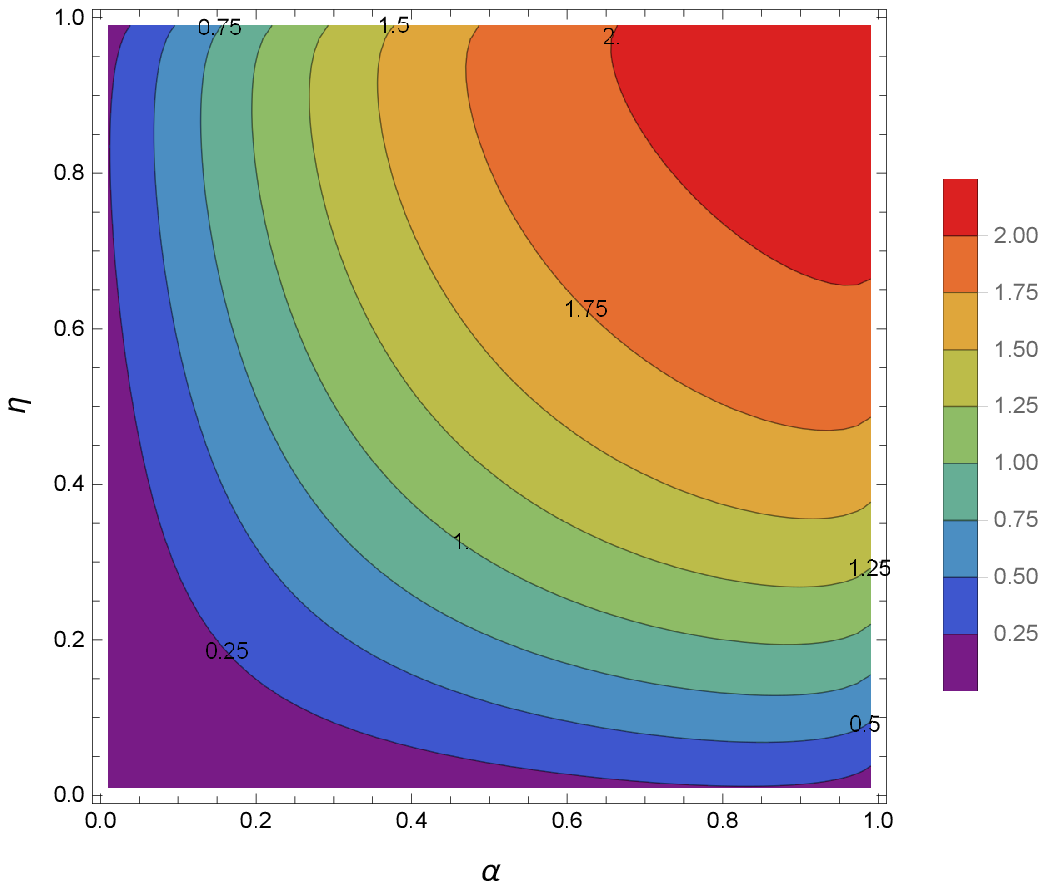}
	\end{minipage} \hfill
	\caption{$\langle M_3 \rangle$ as a function of $\alpha$ and $\eta$ in the second setup~\eqref{BellSetup2} for the squeezed-coherent state~\eqref{Squeezed-Coherent}. In the upper panel, we adopted $\phi=\pi$ and $a=0, \, a'=\pi/2, \, b= - b' = -\pi/4, \, c= - c' = +\pi/4$; in the lower panel, we have $\phi=0$ and $a=0, \, a'=\pi/2, \, b= -b' = -\pi/4, \, c= - c' = +\pi/4$. There is violation whenever we have $\vert \langle M_3 \rangle \vert >2$.}
	\label{M3scAllContour}
\end{figure}

	We plot $ \langle M_3 \rangle $ as a function of $\alpha$, considering $\eta = \alpha + 0.001$ in Fig.~\ref{M3scAll2D}, and exhibit the region of parameters $\alpha$ and $\eta$ leading to the violation of $\vert \langle M_3 \rangle \vert \leq 2$ in Fig.~\ref{M3scAllContour}. We remark that the results obtained for the	squeezed-coherent states in both setups are qualitatively very similar, having only small quantitative differences.

	\vspace{-0.3cm}
	\section{Squeezed-Squeezed states}\label{SecSqueezSqueez}
	\vspace{-0.3cm}
			
	Let us proceed with our study of Mermin's inequality violation considering now 4-partite systems. In order to do so, we define the {\it squeezed-squeezed} entangled state:
	\begin{align}\label{Squeezed-Squeezed}
		\vert \psi \rangle_{SS} = N_{SS} \left( \vert \eta \rangle_{ab} \vert \sigma \rangle_{cd} + e^{i \phi} \vert \sigma \rangle_{ab} \vert \eta \rangle_{cd} \right),
	\end{align}
	where both $\vert \eta \rangle$ and $\vert \sigma \rangle$ represent squeezed states belonging to their respective product Hilbert space as defined in Eq.~\eqref{Squeezed}, with squeezing parameters $\eta$ and $\sigma$. The normalization of this squeezed-squeezed state~\eqref{Squeezed-Squeezed} is
	\begin{align}\label{key}
		N_{SS} = \frac{1}{\sqrt{2}} \left[1 + \cos \phi \, \frac{(1 - \eta^2) (1 - \sigma^2)}{(1 - \eta \sigma)^2}\right]^{-\frac{1}{2}}.
	\end{align}
	
	In the following, we will analyze the Mermin's inequality corresponding to the polynomial $M_4$, adopting the two Bell setups described before, see Eqs.~\eqref{BellSetup1},\eqref{BellSetup2}. As before, in both cases it is sufficient to compute the correlator $\langle ABCD \rangle$, putting primes on the corresponding parameters when appropriate. Thus, using Eq.~\eqref{M4Def} one can construct the correlator $\langle 2 M_4 \rangle$, whose explicit expression will not be reported here to not clutter the text.

\begin{figure}[t!]
	\begin{minipage}[b]{1.0\linewidth}
		\includegraphics[width=\textwidth]{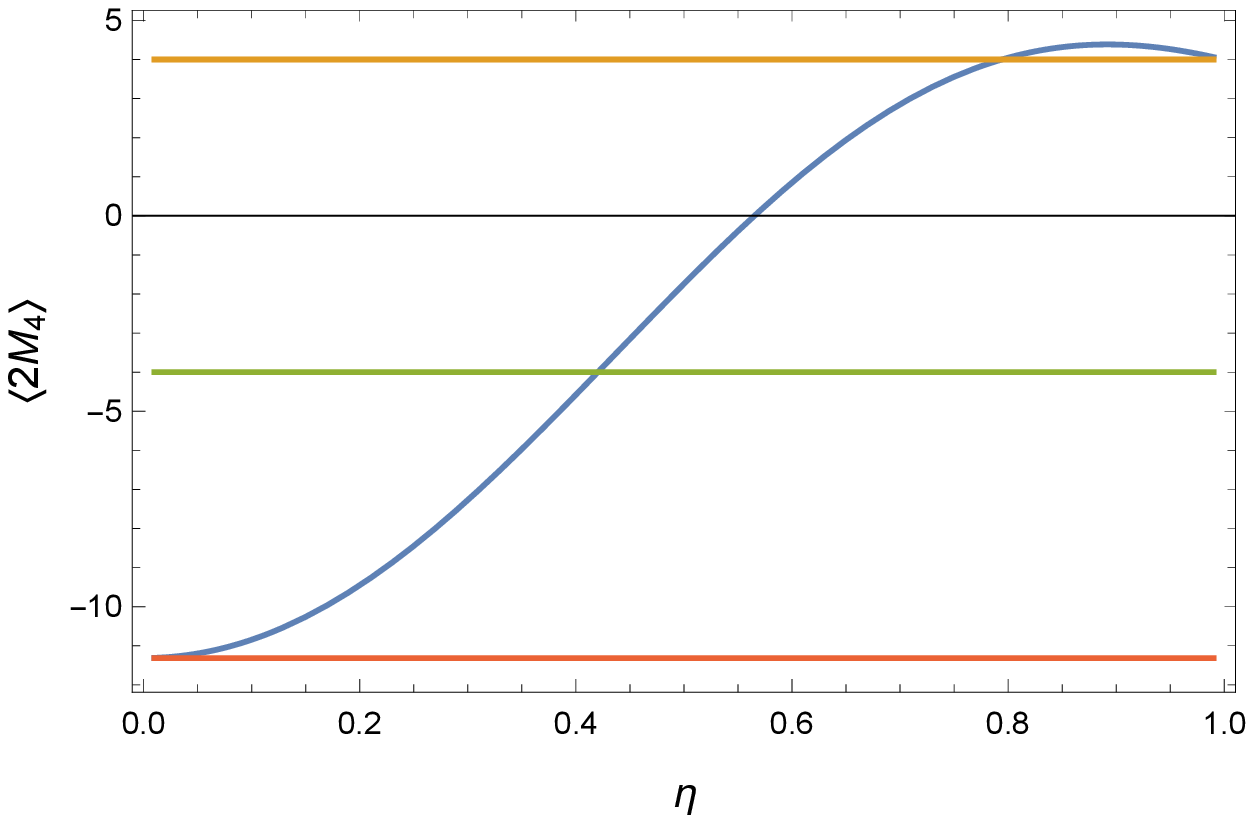}
	\end{minipage} \hfill
	\begin{minipage}[b]{1.0\linewidth}
	\includegraphics[width=\textwidth]{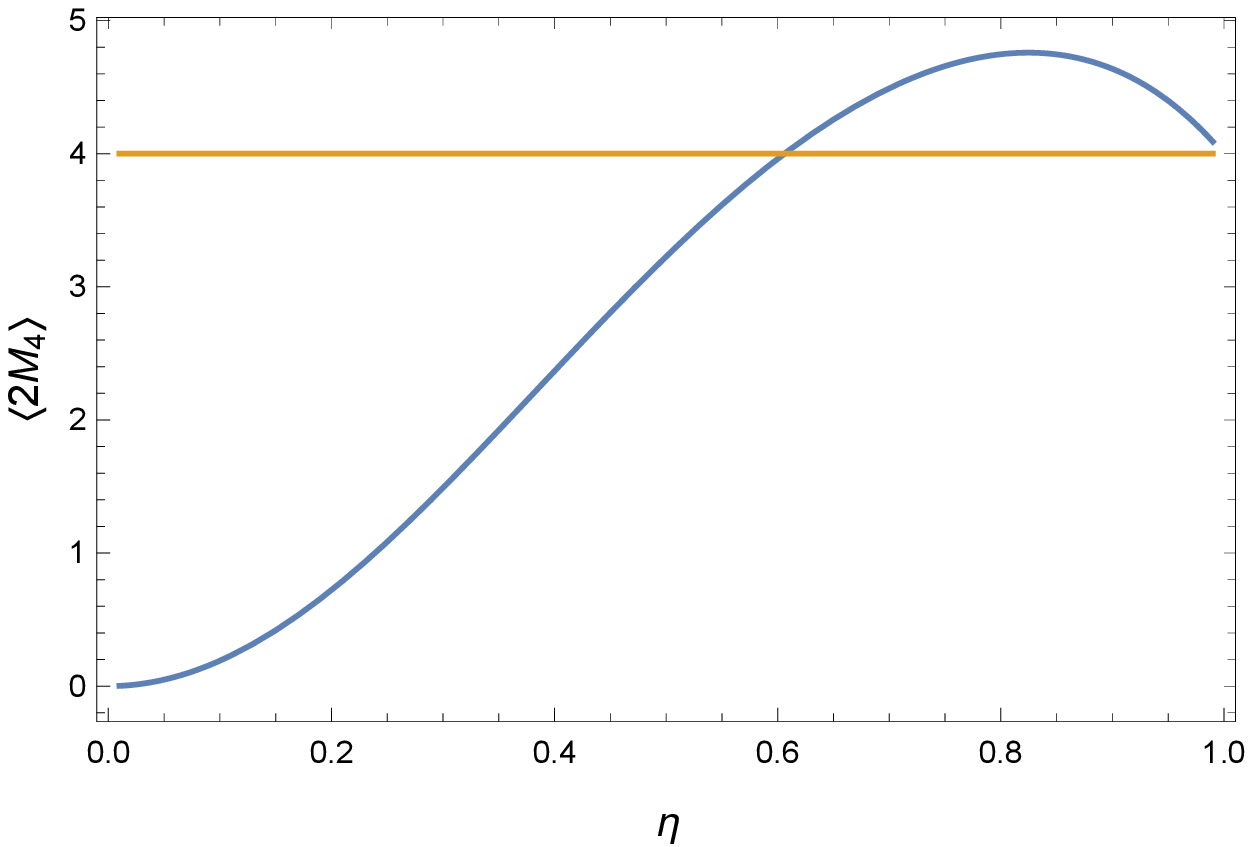}
	\end{minipage} \hfill
	\caption{$\langle 2 M_4 \rangle$ as a function of $\eta$ with $\sigma = \eta + 0.001$ in the first setup~\eqref{BellSetup1} for the squeezed-squeezed state~\eqref{Squeezed-Squeezed}. In the upper panel, we adopted $\phi=\pi$ and $a=0, \, a'=\pi/2, \, b = -b'=-\pi/4, \, c= d  =- c'= -d'= +\pi/4$; in the lower panel, we adopted $\phi=0$ and $a=-\pi/4, \, a'=\pi/2, \, b = -b'=-\pi/4, \, c=-\pi/2, \, c'=0, \, d  =\pi/4, \, d'= 3\pi/4$. There is violation whenever we have $\vert \langle 2 M_4 \rangle \vert >4$. The red line represents the Quantum Mechanics lower bound, given by $\langle 2 M_4 \rangle = -8\sqrt{2}$.}
	\label{M4ssTwo2D}
\end{figure}
\begin{figure}[t!]
	\begin{minipage}[b]{1.0\linewidth}
		\includegraphics[width=\textwidth]{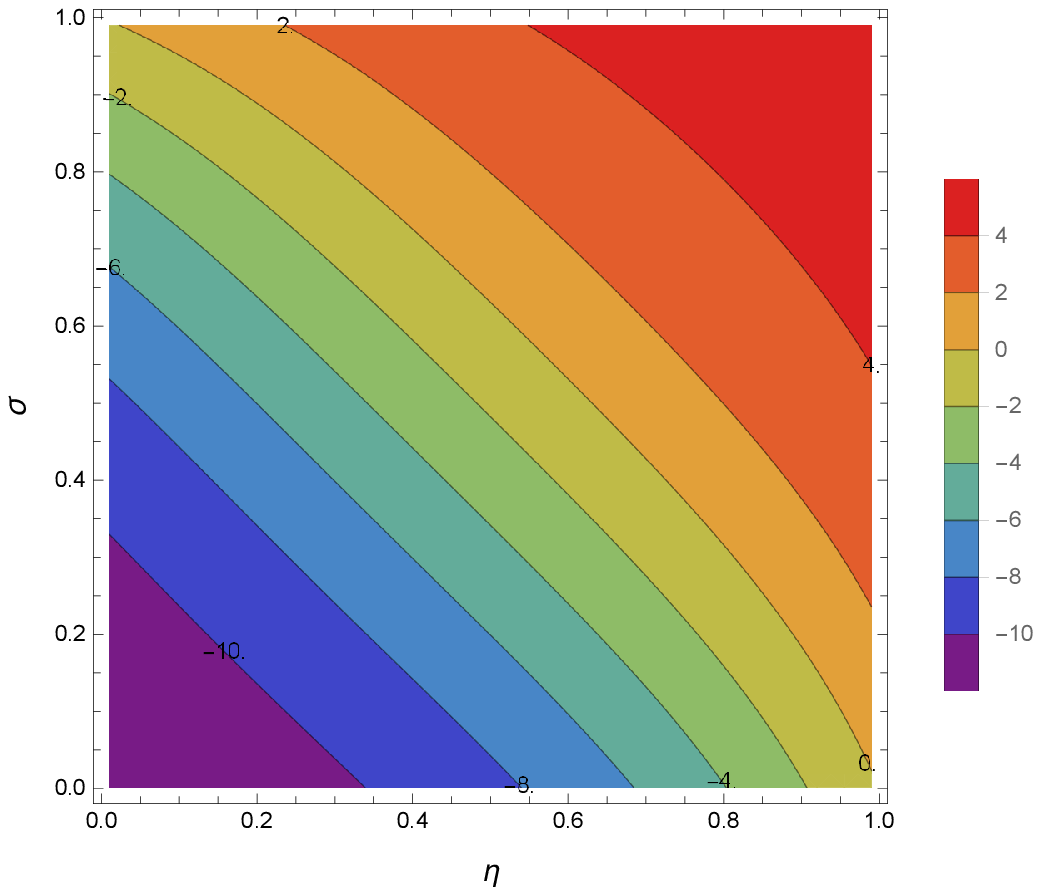}
	\end{minipage} \hfill
	\begin{minipage}[b]{1.0\linewidth}
	\includegraphics[width=\textwidth]{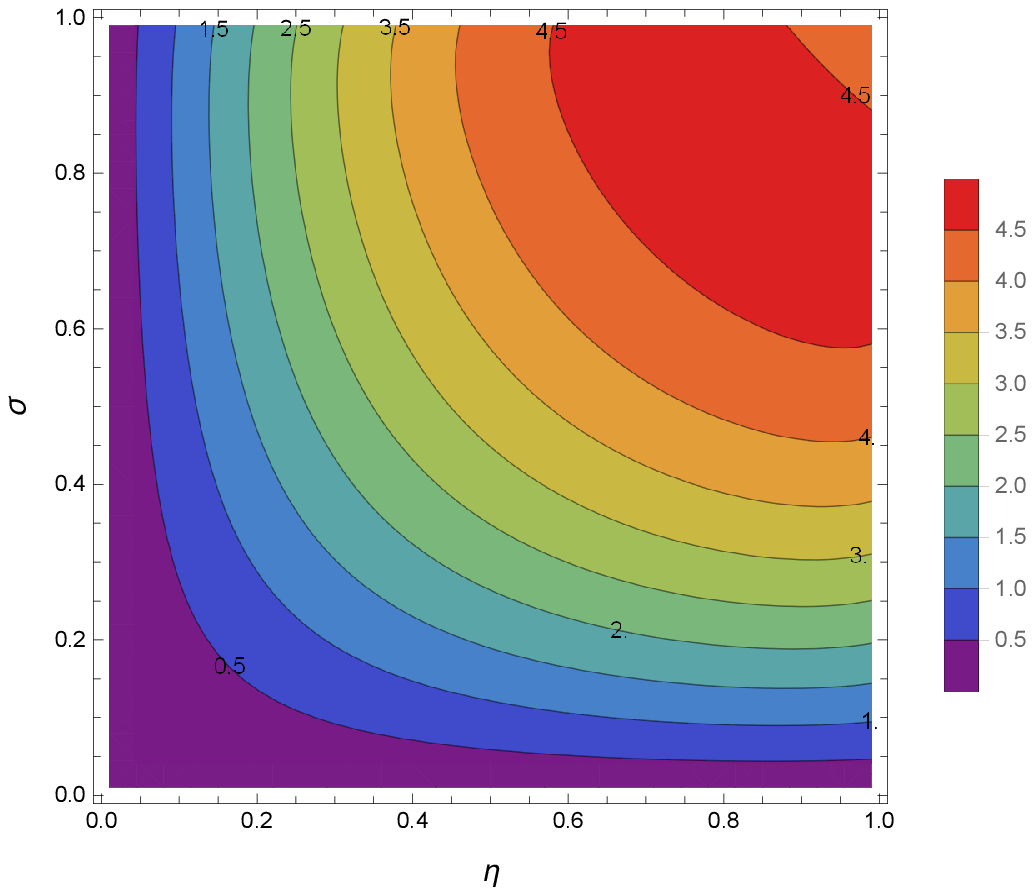}
	\end{minipage} \hfill
	\caption{$\langle 2 M_4 \rangle$ as a function of $\eta$ and $\sigma$ in the first setup~\eqref{BellSetup1} for the squeezed-squeezed state~\eqref{Squeezed-Squeezed}.In the upper panel, we adopted $\phi=\pi$ and $a=0, \, a'=\pi/2, \, b = -b'=-\pi/4, \, c= d  =- c'= -d'= +\pi/4$; in the lower panel, we adopted $\phi=0$ and $a=-\pi/4, \, a'=\pi/2, \, b = -b'=-\pi/4, \, c=-\pi/2, \, c'=0, \, d  =\pi/4, \, d'= 3d$. There is violation whenever we have $\vert \langle 2 M_4 \rangle \vert >4$.}
	\label{M4ssTwoContour}
\end{figure}

	\vspace{-0.2cm}
	\subsection{First setup}
	\vspace{-0.2cm}
	To begin with, we compute the correlator $\langle ABCD \rangle$ considering the squeezed-squeezed state~\eqref{Squeezed-Squeezed}, adopting the first Bell setup defined by Eq.~\eqref{BellSetup1}. We have:
	\begin{align}\label{key}
		\langle AB&CD \rangle = \Omega_{SS} \Big\{ 4 \eta \sigma  \cos(a+b) \cos(c+d) \nonumber \\
		&+ 2 \eta \sigma \cos\phi \cos(a+b+c+d) +   \eta^2 \cos(a+b-c-d+\phi) \nonumber \\
		&+   \sigma^2 \cos(a+b-c-d-\phi) \nonumber \\
		&+  \left(\cos(a+b) + \cos(c+d)\right) \Big[\eta \left(\frac{1}{1-\sigma^2} - 1 - \sigma^2\right) \nonumber \\
		&+ \sigma \left(\frac{1}{1-\eta^2} - 1 - \eta^2\right)\Big] + \left(\frac{1}{1- \sigma \eta} - 1 - \sigma \eta\right) \times  \nonumber \\
		&\times\big[ \eta \left(\cos(a+b+\phi) + \cos(c+d-\phi)\right) \nonumber \\
		&+ \sigma \left(\cos(a+b-\phi) + \cos(c+d+\phi)\right)\big]     \nonumber \\
		&+  \left(\frac{1}{1-\eta^2} - 1 - \eta^2\right) \left(\frac{1}{1- \sigma^2} - 1 - \sigma^2\right)  \nonumber \\
		&+   \cos\phi \left(\frac{1}{1-\eta \sigma} - 1 - \eta \sigma\right)^2  \Big\},  
	\end{align}
	where the overall factor is given by
	\begin{align}\label{OmegaSS}
		\Omega_{SS} =   \frac{(1-\eta^2) (1-\sigma^2)}{\left(1 + \cos \phi \, \frac{(1 - \eta^2) (1 - \sigma^2)}{(1 - \eta \sigma)^2}\right)}.
	\end{align}
	
	With this expression at hand, we can immediately construct the correlator $\langle 2 M_4 \rangle$ and proceed to the numerical analysis, searching for a set of parameters that leads to the violation of Mermin's inequality, {\it i.e.},  $ \vert \langle 2 M_4 \rangle \vert > 4$. 
	
	We were able to find a violation by choosing a relative phase $\phi = \pi$ and adopting the following set of parameters: $a=0, \, a'=\pi/2, \, b = -b'=-\pi/4, \, c= d  =- c'= -d'= +\pi/4$. With these parameters, we can find $\vert \langle 2 M_4 \rangle \vert = 4.4501$ for $(\eta, \sigma) = (0.40,0.41)$, and $\vert \langle 2 M_4 \rangle \vert = 4.4375$ for $(\eta, \sigma) = (0.1,0.7)$, showing that also here we can find violations without taking the parameters close to each other. The Quantum Mechanics bound $\vert \langle 2 M_4 \rangle \vert = 8 \sqrt{2} \approx 11.3137$ is saturated by taking $(\eta, \sigma) = (0.0010, 0.0011)$. We have also found a small violation for $\phi=0$, considering the set of parameters: $a=-\pi/4, \, a'=\pi/2, \, b = -b'=-\pi/4, \, c=-\pi/2, \, c'=0, \, d  =\pi/4, \, d'= 3\pi/4$. In this case, we find  $\vert \langle 2 M_4 \rangle \vert = 4.6220$ for $(\eta, \sigma) = (0.90,0.91)$, and $\vert \langle 2 M_4 \rangle \vert = 4.5460$ for $(\eta, \sigma) = (0.6,0.9)$.
	
	\vspace{-0.1cm}
	The value of $\langle 2 M_4 \rangle$ as a function of $\eta$ with $\sigma = \eta + 0.001$ is shown in Fig.~\ref{M4ssTwo2D}. Notice that for $\phi=\pi$ there is violation for small values of $\eta$, and the Quantum Mechanics bound is saturated for very small values of $\eta$. Remarkably, for this case we can also find a small violation for big values of $\eta$. For $\phi = 0$ we find a small violation for big values of $\eta$. The region of parameters $\eta$ and $\sigma$ leading to the violation of $\vert \langle 2 M_4 \rangle \vert \leq 4$ is exhibited in Fig.~\ref{M4ssTwoContour}.

\begin{figure}[t!]
	\begin{minipage}[b]{1.0\linewidth}
		\includegraphics[width=\textwidth]{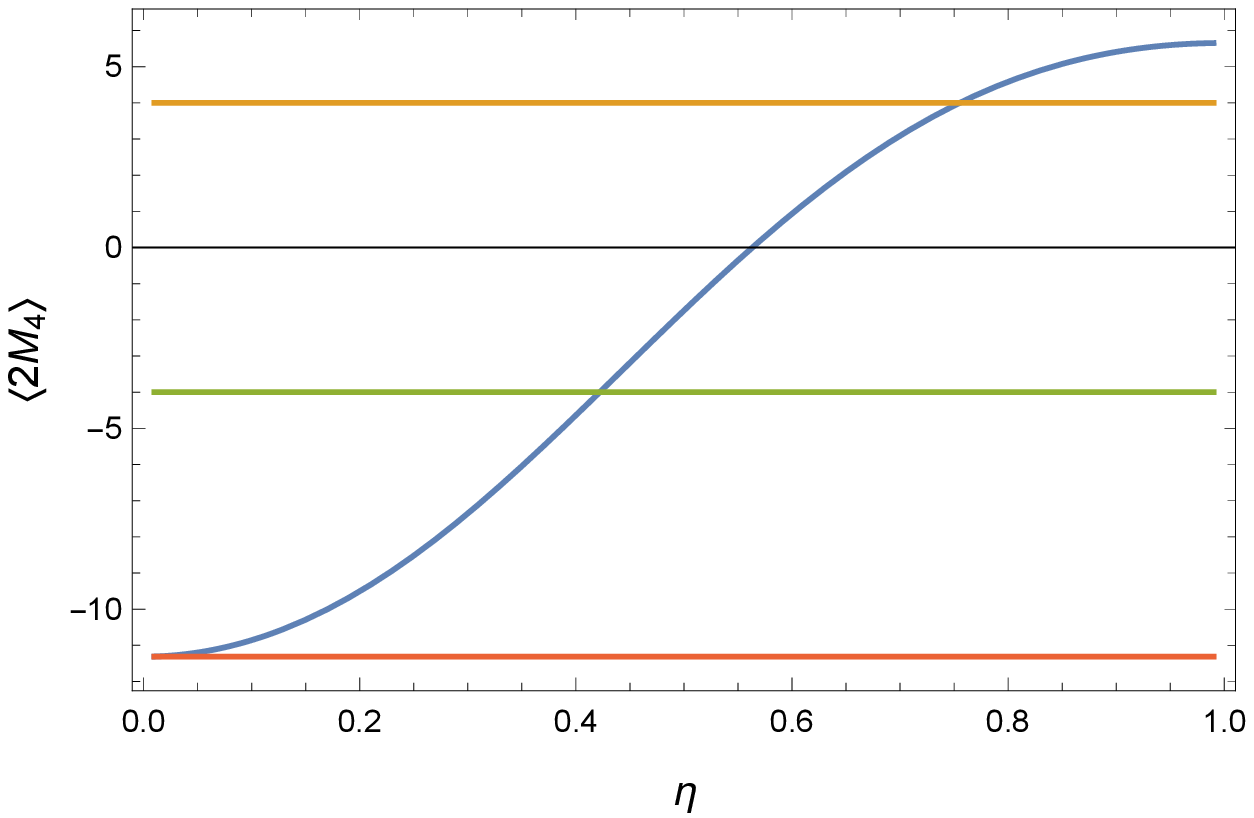}
	\end{minipage} \hfill
	\begin{minipage}[b]{1.0\linewidth}
	\includegraphics[width=\textwidth]{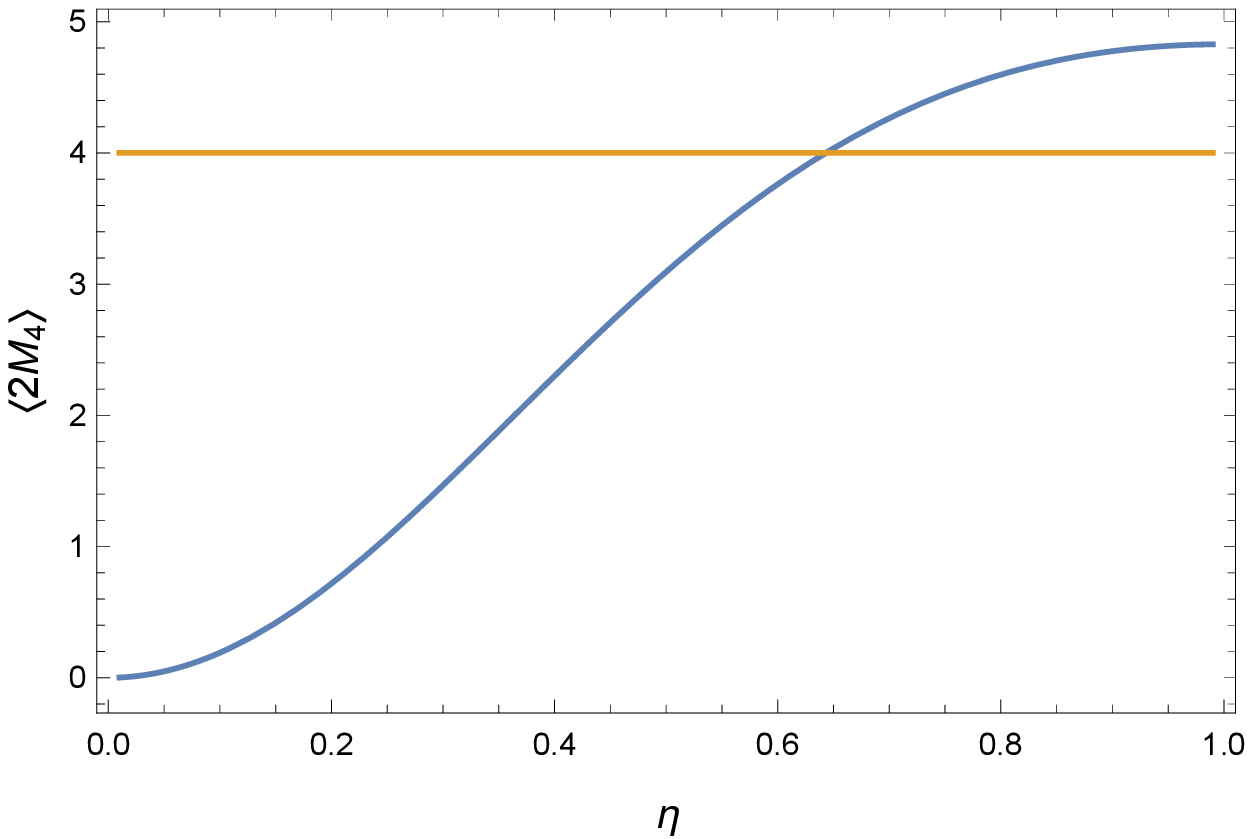}
\end{minipage} \hfill
	\caption{$\langle 2 M_4 \rangle$ as a function of $\eta$ with $\sigma = \eta + 0.001$ in the second setup~\eqref{BellSetup2} for the squeezed-squeezed state~\eqref{Squeezed-Squeezed}.  In the upper panel, we adopted $\phi=\pi$ and $a=0, \, a'=\pi/2, \, b = -b'=-\pi/4, \, c= d  =- c'= -d'= +\pi/4$; in the lower panel, we adopted $\phi=0$ and $a=-\pi/4, \, a'=\pi/2, \, b = -b'=-\pi/4, \, c=-\pi/2, \, c'=0, \, d  =\pi/4, \, d'= 3\pi/4$. There is violation whenever we have $\vert \langle 2 M_4 \rangle \vert >4$. The red line represents the Quantum Mechanics lower bound, given by $\langle 2 M_4 \rangle = -8\sqrt{2}$.}
	\label{M4ssAll2D}
\end{figure}
\begin{figure}[t!]
	\begin{minipage}[b]{1.0\linewidth}
		\includegraphics[width=\textwidth]{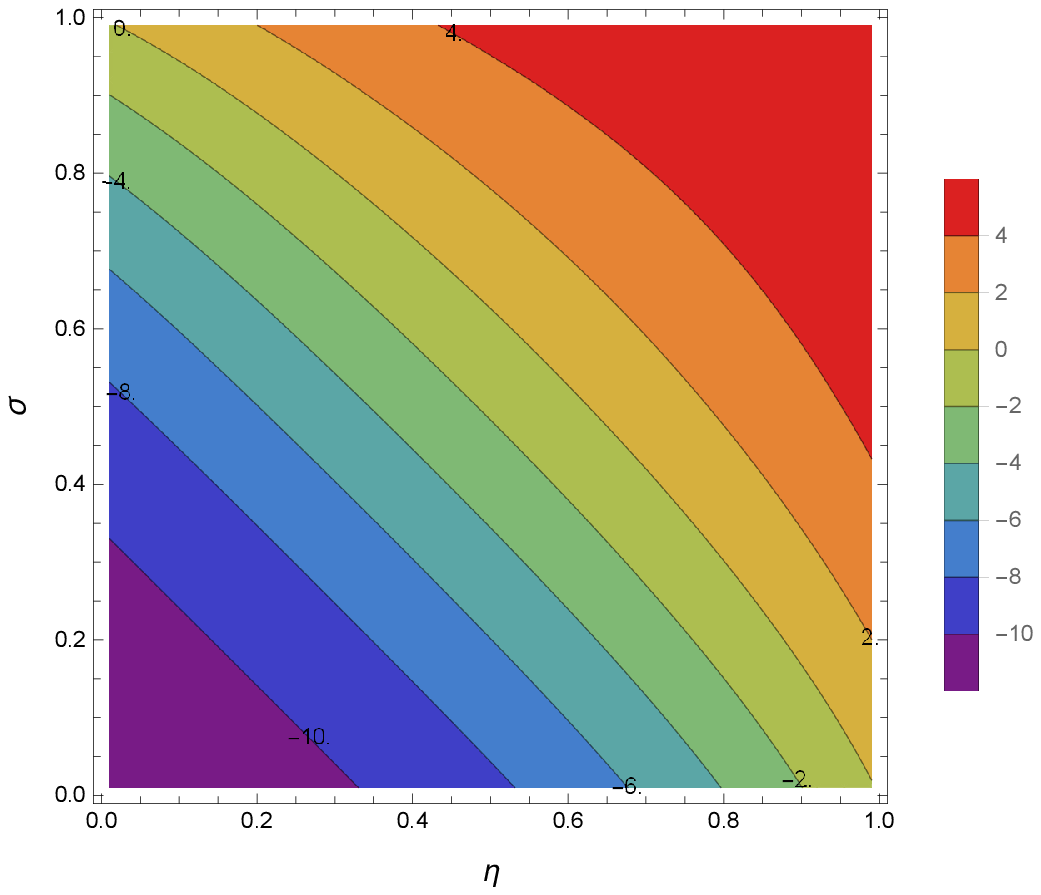}
	\end{minipage} \hfill
	\begin{minipage}[b]{1.0\linewidth}
	\includegraphics[width=\textwidth]{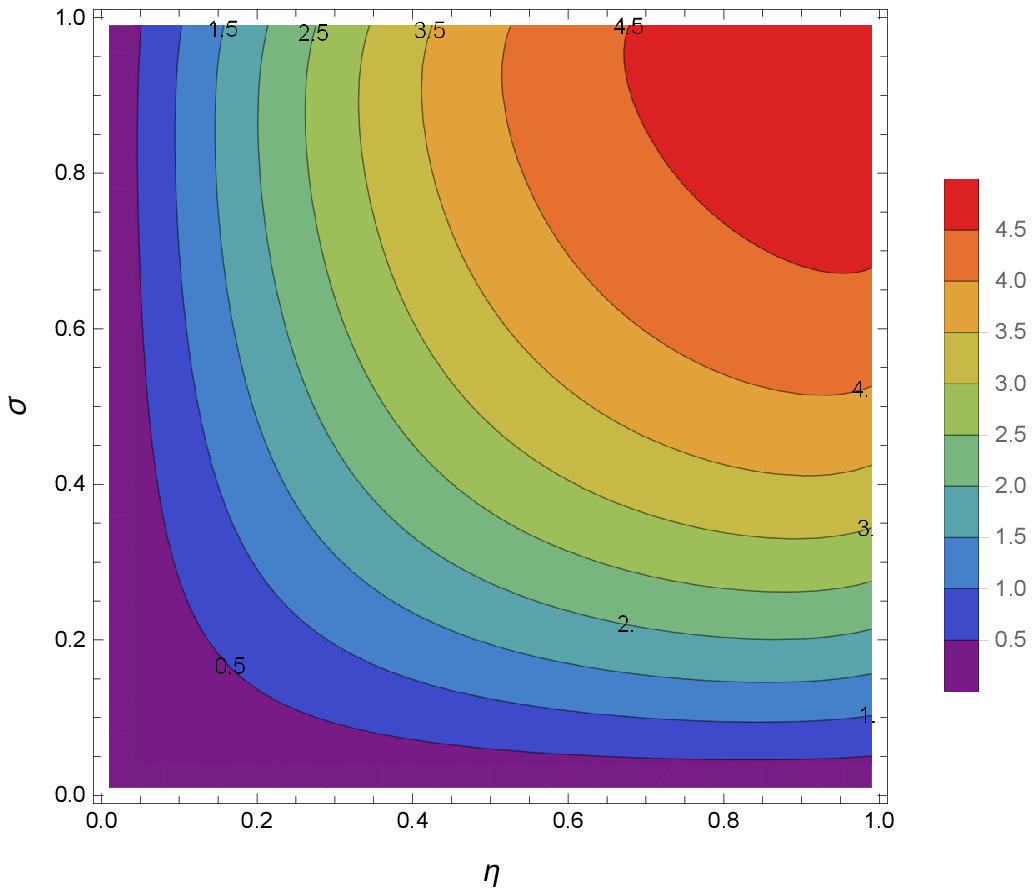}
	\end{minipage} \hfill
	\caption{$\langle 2 M_4 \rangle$ as a function of $\eta$ and $\sigma$ in the second setup~\eqref{BellSetup2} for the squeezed-squeezed state~\eqref{Squeezed-Squeezed}. In the upper panel, we adopted $\phi=\pi$ and $a=0, \, a'=\pi/2, \, b = -b'=-\pi/4, \, c= d  =- c'= -d'= +\pi/4$; in the lower panel, we adopted $\phi=0$ and $a=-\pi/4, \, a'=\pi/2, \, b = -b'=-\pi/4, \, c=-\pi/2, \, c'=0, \, d  =\pi/4, \, d'= 3d$. There is violation whenever we have $\vert \langle 2 M_4 \rangle \vert >4$.}
	\label{M4ssAllContour}
\end{figure}
	\vspace{-0.2cm}
	\subsection{Second setup}
	\vspace{-0.2cm}
	Giving sequence to our analysis, we consider the second setup defined in Eq.~\eqref{BellSetup2}. Thus, the correlator $\langle ABCD \rangle$ in the squeezed-squeezed state~\eqref{Squeezed-Squeezed} is given by
	\begin{align}\label{key}
			&\langle ABCD \rangle = \Omega_{SS} \Big\{  \frac{4 \eta \sigma \cos(a+b) \cos(c+d)}{(1-\eta^4) (1-\sigma^4)} \nonumber \\
			&+ \frac{1}{(1-\eta^2\sigma^2)^2} \Big[2 \eta \sigma \cos \phi \cos(a+b+c+d) \nonumber \\
			&+ \! \eta^2 \! \cos( a \!+\! b \!-\! c \!-\! d \!+\! \phi) + \! \sigma^2 \! \cos( \! a \!+\! b \!-\! c \!-\! d \!-\! \phi)\Big] \Big\},
	\end{align}
	where the overall factor $\Omega_{SS}$ is the same as before~\eqref{OmegaSS}. Notice that we were able to find a closed analytic expression for this case, differently from the squeezed-coherent case in the second setup studied before. 
	
	Once more, we were able to find a violation with a relative phase $\phi = \pi$ and the same set of parameters: $a=0, \, a'=\pi/2, \, b = -b'=-\pi/4, \, c= d  =- c'= -d'= +\pi/4$. Here we obtain $\vert \langle 2 M_4 \rangle \vert = 4.5089$ considering $(\eta, \sigma) = (0.4, 0.41)$ and $\vert \langle 2 M_4 \rangle \vert = 4.4010$ taking $(\eta, \sigma) = (0.1, 0.7)$.  The Quantum Mechanics bound $\vert \langle 2 M_4 \rangle \vert = 8 \sqrt{2}$ is saturated for $(\eta, \sigma) = (0.0010, 0.0011)$. We have also found a violation for $\phi=0$ with the set of parameters: $a=-\pi/4, \, a'=\pi/2, \, b = -b'=-\pi/4, \, c=-\pi/2, \, c'=0, \, d  =\pi/4, \, d'= 3\pi/4$. For instance, we have $\vert \langle 2 M_4 \rangle \vert = 4.7805$ considering $(\eta, \sigma) = (0.90, 0.91)$ and $\vert \langle 2 M_4 \rangle \vert = 4.2955$ taking $(\eta, \sigma) = (0.6, 0.9)$.
	We plot $ \langle 2 M_4 \rangle $ as a function of $\eta$, considering $\eta = \sigma + 0.001$  in Fig.~\ref{M4ssAll2D}. The region of parameters $\eta$ and $\sigma$ leading to the violation of $\vert \langle 2 M_4 \rangle \vert \leq 4$ is shown in Fig.~\ref{M4ssAllContour}. We remark that, as also happened with the squeezed-coherent states, the results obtained for the squeezed-squeezed states in both setups are qualitatively very similar, having only small quantitative differences.

	\vspace{-0.3cm}
	\section{Concluding Remarks}\label{SecConclusion}
	\vspace{-0.3cm}
	In this work, we probed the violation of Mermin's inequalities through two Bell setups built with pseudospin operators~\eqref{pseudospin} for $n$-partite systems, using squeezed-coherent states~\eqref{Squeezed-Coherent} for $n=3$ and squeezed-squeezed states~\eqref{Squeezed-Squeezed} for $n=4$. We have analyzed the set of parameters leading to the explicit violation of the corresponding Mermin's inequalities. 
	Remarkably, we have shown that considering sufficiently small parameters we can not only violate the inequality but also saturate the bound predicted by Quantum Mechanics, thus providing examples of non-GHZ states which lead to a maximal violation. Both setups gave us qualitatively similar results, with small quantitative differences. Interestingly, in the squeezed-squeezed state, besides the large violation with small squeezing parameters, we have also found a small violation with the same relative phase when the parameters are close to 1, something that was not observed in the squeezed-coherent state case. Finally, we remark that the strategy presented here could in principle be generalized for $n$-partite systems with $n>4$, by considering squeezed-coherent states when $n$ is odd and squeezed-squeezed states when $n$ is even.

	\begin{acknowledgments}
		The authors thank the Brazilian agencies CNPq and FAPERJ for financial support. S.~P.~Sorella  and  I.~Roditi are both CNPq researchers, respectively under the contracts 301030/2019-7 and  311876/2021-8.
	\end{acknowledgments}
	


\begin{thebibliography}{99}
		
		
		\bibitem{EPR35}
		A. Einstein, B. Podolsky, and N. Rosen, {\it Can Quantum-Mechanical Description of Physical Reality Be Considered Complete?}, Phys. Rev. {\bf 47}, 777 (1935).
		
		\bibitem{Bell64}
		J. Bell, {\it On the Einstein Podolsky Rosen paradox}, Physics Physique Fizika {\bf 1}, 195 (1964).
		
		\bibitem{CHSH89}
		J. F. Clauser, M. A. Horne, A. Shimony, and R. A. Holt, {\it Proposed Experiment to Test Local Hidden-Variable Theories},	Phys. Rev. Lett. {\bf 23}, 880 (1969).
		
		\bibitem{Aspect82}
		A. Aspect, J. Dalibard and G. Roger, {\it Experimental Test of Bell's Inequalities Using Time-Varying Analyzers}, Phys. Rev. Lett. {\bf 49}, 1804 (1982).
		
		\bibitem{Hensen15}
		B. Hensen et al., {\it Loophole-free Bell inequality violation using electron spins separated by 1.3 kilometres}, Nature {\bf 526}, 682 (2015).
		
		\bibitem{Zou21}
		N. Zou, {\it Quantum Entanglement and Its Application in	Quantum Communication}, J. Phys.: Conf. Ser. {\bf 1827}, 012120 (2021).
		
		\bibitem{Brunner14}
		N. Brunner et al., {\it Bell nonlocality}, Rev. Mod. Phys. {\bf 86}, 419 (2014).
		
		\bibitem{Reid88}
		M. D. Reid and P. D. Drummond, {\it Quantum Correlations of Phase in Nondegenerate Parametric Oscillation}, Phys. Rev. Lett. {\bf 60}, 2731 (1988).
		
		\bibitem{Reid89}
		M. D. Reid, {\it Demonstration of the Einstein-Podolsky-Rosen paradox using nondegenerate parametric amplification}, Phys. Rev. A {\bf 40}, 913 (1989).
		
		\bibitem{Walls94}
		D. F. Walls and G. J. Milburn, {\it Quantum Optics}, (Springer-Verlag, Berlin, 1994).
		
		\bibitem{Ou92a}
		Z. Y. Ou, S. F. Pereira, H. J. Kimble, and K. C. Peng, {\it Realization of the Einstein-Podolsky-Rosen paradox for continuous variables}, Phys. Rev. Lett. {\bf 68}, 3663 (1992).
		
		\bibitem{Ou92b}
		Z.Y. Ou, S. F. Pereira, and H. J. Kimble, {\it Realization of the Einstein-Podolsky-Rosen paradox for continuous variables in nondegenerate parametric amplification}, Appl. Phys. B {\bf 55}, 265 (1992).
		
		\bibitem{Kuzmich00}
		A. Kuzmich, I. A. Walmsley, and L. Mandel, {\it Violation of Bell's Inequality by a Generalized Einstein-Podolsky-Rosen State Using Homodyne Detection}, Phys. Rev.	Lett. {\bf 85}, 1349 (2000).
		
		\bibitem{Banaszek98a}
		K. Banaszek and K. Wódkiewicz, {\it Nonlocality of the Einstein-Podolsky-Rosen state in the Wigner representation}, Phys. Rev. A {\bf 58}, 4345 (1998).
		
		\bibitem{Banaszek98b}
		K. Banaszek and K. Wódkiewicz, {\it Testing Quantum Nonlocality in Phase Space}, Phys. Rev. Lett. {\bf 82}, 2009 (1999).
	
		\bibitem{Grangier88}
		P. Grangier, M. J. Potasek, and B. Yurke, {\it Probing the phase coherence of parametrically generated photon pairs: A new test of Bell's inequalities}, Phys. Rev. A {\bf 38}, 3132 (1988).
		
		\bibitem{Chen02}
		Z-B. Chen, J-W. Pan, G. Hou, and Y-D. Zhang, {\it Maximal Violation of Bell’s Inequalities for Continuous Variable Systems}, Phys. Rev. Lett. {\bf 88}, 040406 (2002).
		
		\bibitem{Tsirelson80}
		B. S. Cirel’son, {\it Quantum generalizations of Bell's inequality
		}, Lett. Math. Phys. {\bf 4}, 93 (1980).
		
		\bibitem{Lloyd99}
		S. Lloyd and S. L. Braunstein, {\it Quantum Computation over Continuous Variables}, Phys. Rev. Lett. {\bf 82}, 1784 (1999).
		
		\bibitem{Vaidman94}
		L. Vaidman, {\it Teleportation of quantum states}, Phys. Rev. A {\bf 49}, 1473 (1994).
		
		\bibitem{Braunstein98a}
		S. L. Braunstein and H. J. Kimble, {\it Teleportation of Continuous Quantum Variables}, Phys. Rev. Lett. {\bf 80}, 869 (1998).
		
		\bibitem{Furusawa98}
		A. Furusawa et al., {\it Unconditional Quantum Teleportation}, Science {\bf 282}, 706 (1998).
		
		\bibitem{Braunstein98b}
		S. L. Braunstein, {\it Error Correction for Continuous Quantum Variables}, Phys. Rev. Lett. {\bf 80}, 4084 (1998).
		
		\bibitem{Lloyd98}
		S. Lloyd and J.-J. E. Slotine, {\it Analog Quantum Error Correction}, Phys. Rev. Lett. {\bf 80}, 4088 (1998).
		
		\bibitem{Braunstein98c}
		S. L. Braunstein, {\it Quantum error correction for communication with linear optics}, Nature {\bf 394}, 47 (1998).
		
		\bibitem{Duan00}
		L-M. Duan et al., {\it Entanglement Purification of Gaussian Continuous Variable Quantum States}, Phys. Rev. Lett. {\bf 84}, 4002 (2000).
		
		\bibitem{Cerf00}
		N. J. Cerf, A. Ipe, and X. Rottenberg, {\it Cloning of Continuous Quantum Variables}, Phys. Rev. Lett. {\bf 85}, 1754 (2000).
		
		
			\bibitem{Greenberger07}
		D. M. Greenberger, M. A. Horne, and A. Zeilinger, {\it Going Beyond Bell's Theorem}, in `Bell's Theorem, Quantum Theory, and Conceptions of the Universe', M. Kafatos (Ed.), Kluwer, Dordrecht, 69-72 (1989), arXiv:0712.0921.
	
		
		\bibitem{Dur00}
		W. D\"ur, G. Vidal, and J. I. Cirac, {\it Three qubits can be entangled in two inequivalent ways}, Phys. Rev. A {\bf 62}, 062314 (2000).
		
		\bibitem{Briegel01}
		H. J. Briegel and R. Raussendorf, {\it Persistent Entanglement	in Arrays of Interacting Particles}, Phys. Rev. Lett. {\bf 86}, 910 (2001).
			
		\bibitem{McCutcheon16}
		W. McCutcheon et al., {\it Experimental verification of multipartite entanglement in quantum networks}, Nat. Commun. {\bf 7}, 13251 (2016).	
			
	
		\bibitem{Bouwmeester99}	
		D. Bouwmeester et al., {\it Observation of Three-Photon Greenberger-Horne-Zeilinger Entanglement}, Phys. Rev. Lett. {\bf 82}, 1345 (1999).
		
		
		\bibitem{Mermin90}
		N. D. Mermin, {\it Extreme quantum entanglement in a superposition of macroscopically distinct states}, Phys. Rev. Lett. {\bf 65}, 1838 (1990).
		
		\bibitem{Ardehali92}
		M. Ardehali, {\it Bell inequalities with a magnitude of violation that grows exponentially with the number of particles}, Phys. Rev. A {\bf 46}, 5375 (1992).
		
		\bibitem{Roy91}
		S. M. Roy and V. Singh, {\it Tests of signal locality and Einstein-Bell locality for multiparticle systems}, Phys. Rev. Lett. {\bf 67}, 2761 (1991); {\it Erratum}, Phys. Rev. Lett. {\bf 70}, 519 (1993).
		
		\bibitem{Belinskii93}
		A. V. Belinskii and D. N. Klyshko, {\it Interference of light and Bell's theorem}, Sov. Phys. Usp. {\bf 36}, 653 (1993).
		
		\bibitem{Gisin98}
		N. Gisin and H. Bechmann-Pasquinucci, {\it Bell inequality, Bell states and maximally entangled states for n qubits}, Phys. Lett. A {\bf 246}, 1 (1998).
		
		\bibitem{GHZ90}
		D. M. Greenberger, M. A. Horne, A. Shimony, and A. Zeilinger, {\it Bell’s theorem without inequalities}, Am. J. Phys. {\bf 58}, 1131 (1990).
		
		\bibitem{Swain18}
		M. Swain, A. Rai, B. K. Behera, and P. K. Panigrahi, {\it Experimental demonstration of the violations of Mermin's and Svetlichny's inequalities for W- and GHZ-class of states}, Quantum Inf. Process {\bf 18}, 218 (2019). 
	
		\bibitem{Werner01}
		R. F. Werner and M. M. Wolf, {\it All-multipartite Bell-correlation inequalities for two dichotomic observables per site}, Phys. Rev. A {\bf 64}, 032112 (2001).
		
		\bibitem{Cereceda01}
		J. L. Cereceda, {\it Mermin's n-particle Bell inequality and operators' noncommutativity}, Phys. Lett. A {\bf 286}, 376 (2001).
		
		\bibitem{Zukowski02}
		M. Zukowski and C. Brukner, {\it Bell’s Theorem for General N-Qubit States}, Phys. Rev. Lett. {\bf 88}, 210401 (2002).		
		
		\bibitem{Collins02b}
		D. Collins et al., {\it Bell-Type Inequalities to Detect True $n$-Body Nonseparability}, Phys. Rev. Lett. {\bf 88}, 170405 (2002).
		
		\bibitem{Pan00}
		J-W. Pan et al., {\it Experimental test of quantum nonlocality in three-photon Greenberger-Horne-Zeilinger entanglement}, Nature 403, 515 (2000).

		\bibitem{Erwen14}
		C. Erven et al., {\it Experimental three-photon quantum nonlocality under strict locality conditions}, Nature Photonics {\bf 8}, 292 (2014).
		
		\bibitem{Zhao03}
		Z. Zhao et al., {\it Experimental Violation of Local Realism by Four-Photon Greenberger-Horne-Zeilinger Entanglement}, Phys. Rev. Lett. {\bf 91}, 180401 (2003).
		
		\bibitem{Ansmann09}
		M. Ansmann et al., {\it Violation of Bell's inequality in Josephson phase qubits}, Nature  {\bf 461}, 504 (2009).
		
		\bibitem{DiCarlo10}
		L. Di Carlo et al., {\it Preparation and measurement of three-qubit entanglement in a superconducting circuit}, Nature {\bf 467}, 574 (2010).
		
		\bibitem{Neeley10}
		M. Neeley et al., {\it Generation of three-qubit entangled states using superconducting phase qubits}, Nature {\bf 467}, 570 (2010).	
		
		\bibitem{Storz23}
		S. Storz et al., {\it Loophole-free Bell inequality violation with superconducting circuits}, Nature {\bf 617}, 265 (2023).
		
		\bibitem{Alsina16a}
		D. Alsina and J. I. Latorre, {\it Experimental test of Mermin inequalities on a five-qubit quantum computer}, Phys. Rev. A {\bf 94}, 012314 (2016).
		
		\bibitem{Alsina16b}
		D. Alsina et al., {\it Operational approach to Bell inequalities: Application to qutrits}, Phys. Rev. A {\bf 94}, 032102 (2016).
			
		\bibitem{Zhang02}
		Z-B. Chen and Y-D. Zhang, {\it Greenberger-Horne-Zeilinger nonlocality for continuous variable systems}, Phys. Rev. A {\bf 65}, 044102 (2002). 
		
		\bibitem{Dorantes09}
		M. M. Dorantes and J. L. Lucio M., {\it Generalizations of the pseudospin operator to test the Bell inequality for the TMSV state}, J. Phys. A: Math. Theor. {\bf 42}, 285309 (2009).
		
		\bibitem{Larsson03}
		J-A. Larsson, {\it Qubits from number states and Bell inequalities for number measurements}, Phys. Rev. A {\bf 67}, 022108 (2003).	
			
		\bibitem{BellCoherent}
		P. De Fabritiis, F. M. Guedes, G. Peruzzo, and S.~P.~Sorella, {\it Entangled coherent states and violations of Bell-CHSH inequalities}, arXiv:2305.04674
		
		
		\bibitem{MerminQFT}
		P. De Fabritiis,  I. Roditi, and S. P. Sorella, {\it Mermin’s inequalities in Quantum Field Theory}, arXiv:2303.12195
	
		\bibitem{Gisin92} 
		N. Gisin and A. Peres, {\it Maximal violation of Bell's inequality for arbitrarily large spin}, Phys. Lett. A {\bf 162}, 15 (1992).
	
		\bibitem{Peruzzo23}
		G. Peruzzo and S. P. Sorella, {\it Entanglement and maximal violation of the CHSH inequality in a system of two spins j: A novel construction and further observations}, Phys. Lett. A \textbf{474}, 128847 (2023).
	
		\bibitem{Sorella23a}
		S. P. Sorella, {\it A study of the violation of the Bell-CHSH inequality}, arXiv:2302.02385.
		
		\bibitem{Sorella23b}
		S. P. Sorella, {\it On the Representations of Bell’s Operators in Quantum Mechanics},  Found. Phys. {\bf 53}, 59 (2023).
		
		
	
		
	
				
	\end{thebibliography}
\end{document}